\documentstyle[epsf,aas2pp4]{article}
\slugcomment{Submitted to ApJS}
\lefthead{Massa and Fitzpatrick}
\righthead{Recalibration of {\it IUE}}

\begin{document}

\title
{A recalibration of {\it IUE}\/ NEWSIPS low dispersion data}

\author
{D. Massa}

\affil
{Raytheon STX, NASA/GSFC, Mailstop 631.0, Greenbelt, MD 20771; 
massa@xfiles.gsfc.nasa.gov}

\and

\author
{E. L. Fitzpatrick}

\affil
{Astronomy Department, Villanova, Villanova, PA 19085; fitz@ast.vill.edu}

\begin{abstract}
While the low dispersion {\it IUE}\/ NEWSIPS data products represent a
significant improvement over original {\it IUE}\/ SIPS data, they still
contain serious systematic effects which compromise their utility for
certain applications.  We show that NEWSIPS low resolution data are
internally consistent to only 10-15\% at best, with the majority of the
problem due to time dependent systematic effects.  In addition, the
NEWSIPS flux calibration is shown to be inconsistent by nearly 10\%.

We examine the origin of these problems and proceed to formulate and
apply algorithms to correct them to $\sim 3\%$ level -- a factor of 5
improvement in accuracy.  Because of the temporal systematics, transforming
the corrected data to the {\it IUE}\/ flux calibration becomes ambiguous.
Therefore, we elect to transform the corrected data onto the {\it HST}\/
FOS system.  This system is far more self-consistent, and transforming
the {\it IUE}\/ data to it places data from both telescopes on a single
system.

Finally, we argue that much of the remaining 3\% systematic effects in the
corrected data is traceable to problems with the NEWSIPS intensity
transformation function (ITF).  The accuracy could probably be doubled by
rederiving the ITF.
\end{abstract}


\section{Introduction}\label{intro}

This paper addresses problems with the absolute flux calibrations, 
thermal corrections, and time-dependent sensitivity corrections of the 
{\it IUE}\/ (Boggess et al.\ 1978) low-dispersion NEWSIPS (``final 
archive'') data (Nichols \& Linsky 1996, NL hereafter). We demonstrate that
there are systematic errors of up to 15 \% in NEWSIPS fluxes and describe 
correction procedures which reduce the systematics to a level compatible 
with the best possible signal-to-noise ratio achievable by a single 
{\it IUE}\/ spectrum, i.e, $\sim3$\%.

While carrying out an independent research program which involved 
fitting Kurucz (1991) model atmospheres to NEWSIPS low-dispersion data
of B-type stars (Fitzpatrick \& Massa 1999), it became apparent that
the absolute flux calibration was suspect. The basis of our suspicion 
was that the model fit residuals were large, strongly wavelength 
dependent, {\em and independent of stellar spectral type}. Since we 
were fitting energy distributions of main sequence stars throughout the
range $10,000 < T_{eff} < 30,000$ K, it was difficult to imagine a 
single ionic signature that would produce such an effect.  
Consequently, we performed a detailed assessment of the NEWSIPS 
low-dispersion data, ultimately involving more than 4600 spectra.  This 
investigation revealed that not only is the absolute flux calibration 
of the NEWSIPS data inconsistent with its proposed standard, but that 
the data also contain serious thermal and time-dependent systematics.

\begin{figure}[t]
\centerline{\hbox{
\epsfxsize=3.5truein
\epsffile{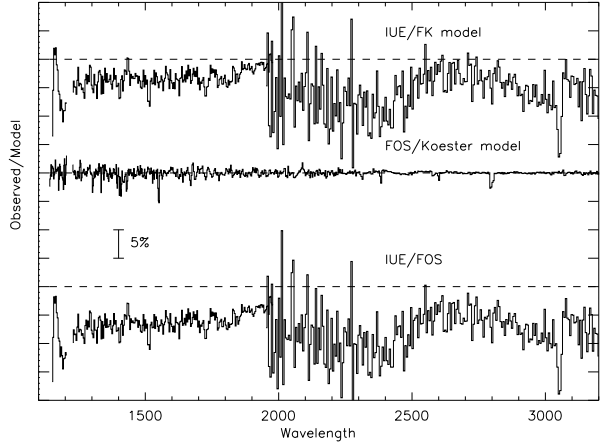}
}}
\caption{Ratios of white dwarf models and observations.  The top panel
shows the ratio of the mean {\it IUE}\/ NEWSIPS data for G191B2B divided
by the Finley \& Koester model used to define the NEWSIPS calibration.
The middle panel if the ratio of the {\it HST}\/ FOS observations of G191B2B
and the Koester model used in the calibration of the FOS.  The bottom panel
is the ratio of the NEWSIPS data divided by the FOS data.}
\label{ratios0}
\end{figure}

The nature of the NEWSIPS absolute flux calibration problem is illustrated
in Figure \ref{ratios0}, which compares the available {\it IUE}\/ NEWSIPS
SWP and LWP data (see \S \ref{data}), {\it HST}\/ FOS data, and models for
the hot white dwarf G191B2B.  The top spectrum is a ratio of the {\it
IUE}\/ NEWSIPS data and the Finley \& Koester (FK) model used to define
the {\it IUE}\/ calibration (NL).  There is obvious disagreement between
the model and the NEWSIPS data, averaging $\sim 5$\% and as large as 10\%
in the LWP. Furthermore, the wavelength-dependent structure in the ratio
is nearly identical to the $T_{eff}$-independent residuals we observed in
our B star fits --- including the high frequency ``noise'' visible in the
figure --- indicating that this results from systematic calibration errors
and {\em not} from random noise.  The middle plot is the ratio of the FOS
fluxes of G191B2B divided by the model provided by Koester to calibrate
the FOS (Bohlin 1996, B96 hereafter).   In this case, the agreement is
excellent, but not exact, since G191B2B  was only one of 8 white dwarfs
used to calibrate the FOS (B96).  The bottom plot is a ratio of the
NEWSIPS and FOS data.  It is very similar to the top ratio, but is also
displaced downward by a wavelength independent, gray, displacement of
$\sim$ 2-3\%.  The displacement results from the {\it IUE}\/ and FOS
projects adopting slightly different models and different scalings for
their UV calibrations; FOS is based on optical photometry calibrations
(B96), and {\it IUE}\/ on UV calibrations (NL).

\begin{figure}[h]
\centerline{\hbox{
\epsfxsize=3.5truein
\epsffile{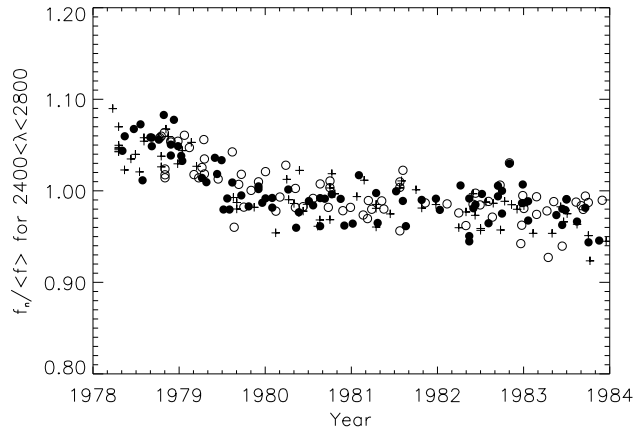}
}}
\caption{Time dependence of standard star fluxes in the LWR camera.
This figure shows the mean flux over the wavelength band $2400< \lambda <
2800$ \AA\/ from normalized by its mean value for 3 {\it IUE}\ standard
stars; HD 60753, BD$+28^\circ4211$ and BD$+75^\circ375$.  The mean flux
was determined for each standard star observation then divided by the
sample mean for that star.  The 3 sets of relative fluxes were then
overploted, with crosses for HD 60753, filled circles for BD$+28^\circ
4211$, and open circles for BD$+75^\circ 375$.}
\label{tlwr}
\end{figure}

An example of the systematic time-dependent effects present in NEWSIPS
data is illustrated in Figure \ref{tlwr} for the {\em IUE}\/ LWR camera.
This plot shows the mean LWR flux in the wavelength band $2400< \lambda <
2800$ \AA\ for 3 standard stars as a function of time during the years
when the LWR was the  default long wavelength camera.  The data for each
star were normalized by  the overall mean for that star during the entire
time interval.  Notice that  each star shows the same time-dependence,
indicating that the effect is instrumental, and not intrinsic to a
particular object.  The magnitude of the effect over this period is
roughly 10\%.  The LWR camera shows the largest level of time-dependence. 
A preliminary discussion of the time dependent behavior of both the
NEWSIPS low {\em and} high dispersion data can be found in Massa et al.
(1998).

\begin{figure}[h]
\centerline{\hbox{
\epsfxsize=5truein
\epsffile{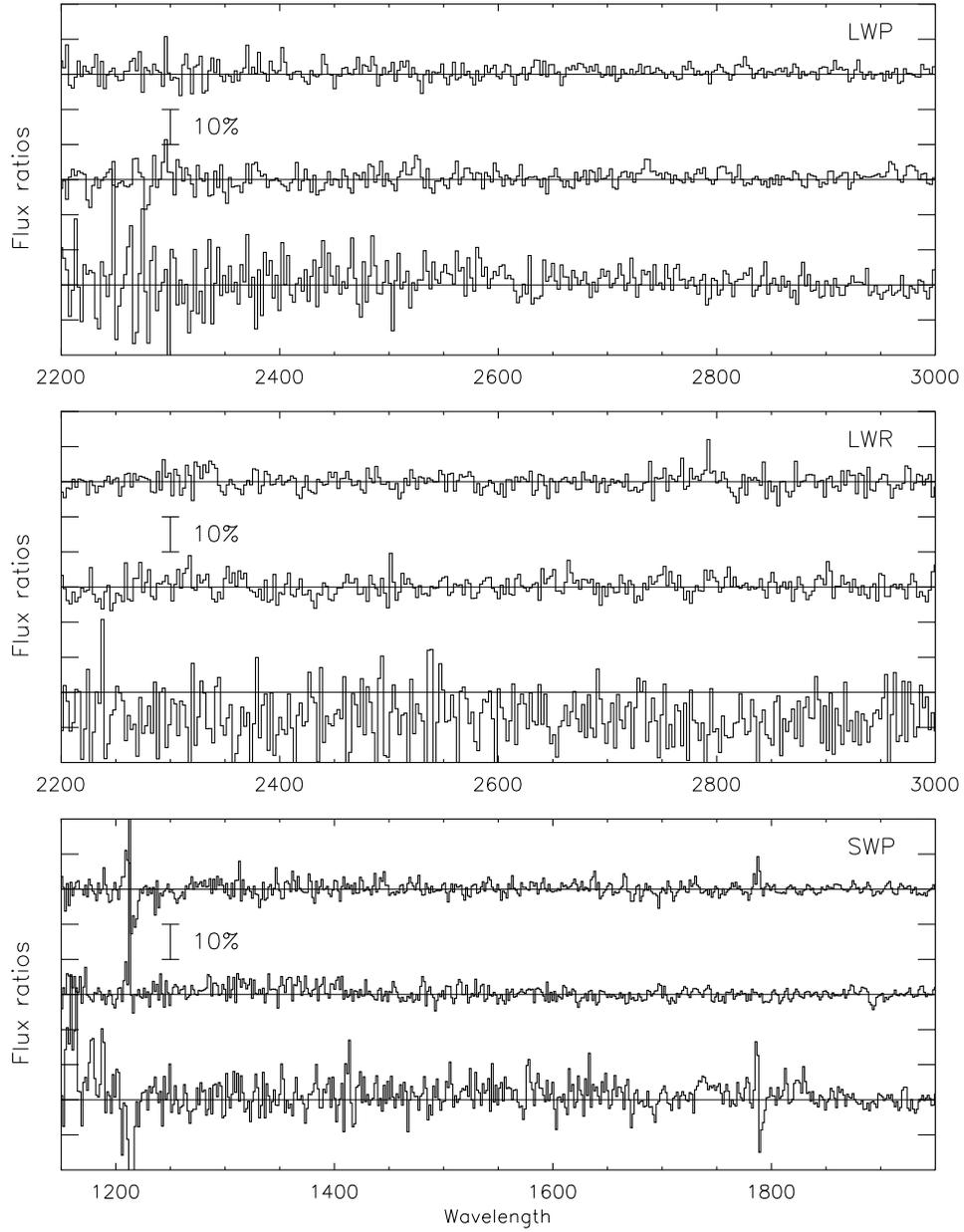}
}}
\caption{This figure demonstrates the presence of high frequency (in
wavelength) time dependent systematics in the {\it IUE}\/ NEWSIPS data (see
\S \ref{intro}).}
\label{highfreq}
\end{figure}

High frequency (``noise-like'') temporal systematics are also present in
the data.  Figure \ref{highfreq} illustrates these effects.  We first
normalize the large aperture spectra of 3 {\it IUE}\/ standard stars (HD
60753, BD$+28^\circ4211$ and  BD$+75^\circ375$ -- see \S 3) by their mean
spectra.  The normalized spectra  were then arranged chronologically.  The
first 40 spectra of this set were divided into 2 20--spectrum samples --
each composed of every other spectrum.  This insures that the 20--spectrum
samples have effectively identical temporal distributions, so time
dependent systematics affect their means in the same way.  The top curve 
in each panel is a ratio of the 2 20--spectrum means.  The middle curve in 
each panel is a ratio of 2 20--spectrum means obtained from the last 40 
spectra of the same 3 star sample and similarly prepared.  Finally, the 
bottom plot in each panel is a ratio of the 2 40--spectrum means obtained 
from all of the spectra used in the first and second curves.  While time 
dependent effects will cancel in the first 2 curves, they are maximized 
in the last one, since it compares data obtained at the beginning and end 
of the mission.  If the high frequency structure was due to random noise, 
its amplitude in the bottom curve would be $1/\sqrt{2}$ times smaller than 
in the top two curves, since twice as many spectra are used to produce the 
ratio.  However, the amplitude of the high frequency structure is clearly 
larger in the bottom ratio!  This is because much of the structure is not 
due to random noise, but to time dependent effects in the data.  Figure
\ref{highfreq} demonstrates that the peak-to-peak amplitude of the high
frequency systematics often exceeds 10\%.

\begin{figure}[h]
\figurenum{4a}
\centerline{\hbox{
\epsfxsize=4.5truein
\epsffile{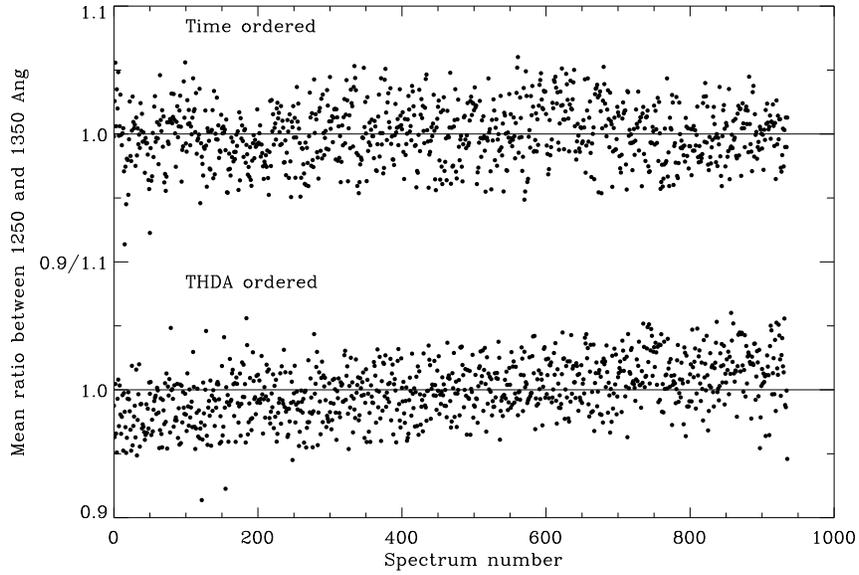}
}}
\caption{Normalized SWP fluxes of {\it IUE}\/ standard stars binned over 
the wavelength band $1250 \leq \lambda \leq 1350$.  The upper plot shows 
the data arranged chronologically, and the lower plot shows them arranged 
in order of ascending $T\!H\!D\!A$ value.}
\label{thdasys}
\end{figure}

In addition to temporal effects, residual temperature effects are also
present in the NEWSIPS data.  These are characterized by the camera 
head amplifier temperature, $T\!H\!D\!A$, measured at the beginning of the 
exposure (these values are supplied as part of the NEWSIPS headers, see, 
Garhart et al., 1997).  The $T\!H\!D\!A$ effects are localized in 
wavelength.  To demonstrate the effect, we used the same data set described 
in the previous paragraph, and binned them over a small wavelength band.  
These data are then plotted in chronological order in the top portions of 
Figures \ref{thdasys} and b.  The same data are then rearranged by 
$T\!H\!D\!A$  value and plotted in the lower portions of the figures.  The 
presence of a systematic $T\!H\!D\!A$ effect on the order of 5\% is 
obvious.  We only show the LWP and SWP data, since the large temporal 
effects in the LWR data tend to obscure the smaller $T\!H\!D\!A$ 
systematics.

The previous examples demonstrate that the {\it IUE}\/ absolute flux
calibration is inconsistent with its reference model by as much as 10\%,
that the LWR data contain time-dependent errors of similar magnitudes,
that all of the data contain high frequency temporal effects whose
amplitudes exceed 10\%, and that thermally induced systematics on the order
of 5\% are also present.  We emphasize that systematic errors of this
order can be important in many applications and must be corrected for the
following reasons:

\begin{enumerate} \item  To utilize the unprecedented temporal baseline of
{\it IUE}, time dependent instrumental drifts in the data must be
corrected to the level of  the maximum achievable $S/N$.

\item  To take full advantage of the signal-to-noise ($S/N$) capabilities
of {\it IUE}, an absolute calibration whose uncertainty is equal to or
less than the maximum achievable $S/N$ is required.

\item  Time dependent dependent high frequency structure can mimic the 
strengthening or weakening of spectral features.  These effects can
produce misleading interpretations and act to nullify potential noise
reductions gained by averaging spectra.  They must be reduced to the level
of typical point-to-point noise.

\end{enumerate}

So what is the maximum achievable $S/N$ for {\it IUE}\/ data?  In spite of
the limited dynamic range of the {\it IUE}\/ detectors, we demonstrate in
\S 3 that a $S/N \sim 30\!:\!1$ should, in principle, be possible for a
single spectral resolution element in an optimally exposed spectrum. 
Furthermore, many bright objects were either observed more than once or
else observed in the trail mode, which could increase the $S/N$ by a
factor 2 (see \S 3). As a result, a large fraction of the objects in the
{\it IUE}\/ archive have a potential $S/N \sim 30\!:\!1$ -- roughly 4 - 5
times more accurate than the current NEWSIPS calibrations.

In the following sections we examine more closely the systematic effects
present in {\it IUE}\/ NEWSIPS data and present a scheme for the removal
of these effects and a correction to the absolute calibration.   Bless and
Percival (1998) performed a critical review of the available UV
calibrations and deduced that the FOS absolute calibration is superior. 
Consequently, we  elected to derive a transformation between the {\it
IUE}\/ NEWSIPS and FOS  systems, rather than recalibrating {\it IUE} using
the G191B2B model.  This also insures that both datasets are on a common
scale.

In \S \ref{stars}, we introduce the stars to be used in our analysis and
indicate which ones will be used to determine temporal and thermal trends,
to derive an absolute flux calibration, and to verify the results. In \S
\ref{data}, we describe some basic characteristics of {\it IUE}\/ and 
explain the different observing modes and why each must be calibrated 
separately. We also discuss how the available data were culled into our 
final sample. In \S \ref{anal}, we describe the mathematical formulation of 
how we correct for systematic effects. In \S \ref{results} we apply the 
analysis to the {\it IUE}\/ data and present our results.  In \S 
\ref{verify}, we verify our results by applying them to sequences of 
spectra not included in the derivation of the corrections.  In \S 
\ref{errors}, we analyze both the random and systematic errors present in 
the corrected data.  In \S \ref{summary}, we summarize our conclusions and 
discuss the availability of IDL programs which apply the corrections to 
NEWSIPS low dispersion data.

\section{The program stars}\label{stars}

Basic data for the program stars used in this study are listed in Table
1.  The stars were selected from the {\it IUE}\/ standard stars compiled
by P\'{e}rez et al.\ (1990) and from the FOS calibration stars given
in B96.  We also indicate in the table the role of each star in our
analysis.  The  ``temporal/thermal standards'' are used derive the time-
and temperature- dependence of the instrumental response, and the
``temporal/thermal control'' stars are used to verify the results. The
``flux standards'' are the stars used to derive the transformation between
the {\it IUE}\/ NEWSIPS and FOS flux scales and the ``flux control'' stars
are used to verify these results.

\section{The data}\label{data}

In this section, we describe the selection of the spectra used in our
analysis.  We begin with a discussion of a few general characteristics of
{\it IUE}\/ data and observations (\S 3.1) and then describe the 4 
{\it IUE}\/ observing modes (\S 3.2).  Lastly, we list the criteria used to 
select or reject individual spectra and individual data points (\S 3.3).  
For a more detailed description of the general properties of {\it IUE}\/ 
and its data, see Newmark et al.\ (1992) and Garhart et al.\ (1997).

\subsection{Overview}

{\it IUE}\/ had two UV spectrographs, covering the short and long
wavelengths regions.  Each spectrograph could send its output to either of
2 (primary and redundant) cameras.  Consequently, {\it IUE}\/ spectra are
referred to as long wavelength primary (LWP), long wavelength redundant
(LWR) or short wavelength prime (SWP).  The short wavelength redundant
camera never operated properly and there are no NEWSIPS data for it. The
wavelength coverages of the spectrograph/camera combinations were $1150 <
\lambda < 1975$ \AA\ for the SWP, $1910 < \lambda <3300$ \AA\ for the LWP
and $1860 < \lambda <3300$ \AA\ for the LWR.

Each spectrograph could be accessed through one of 2 object apertures. One
was a $10\times 20$ arc sec oval, the large aperture, and the other was a
3 arc sec circle, the small aperture.  The size of the telescope image in
the aperture plane was $\sim 4$ arc sec at best, and thus overfilled the
small aperture by a large margin (see Garhart et al. for a more detailed
discussion).

The {\it IUE}\/ detectors had $768\times 768$ pixels and each pixel had a
very limited dynamic range, with only 256 distinct output levels (8 bits).
Since {\it IUE}\/ was in high earth orbit, part of the output signal was
due to particle background, as well as to a pedestal of read noise. 
Consequently, the maximum $S/N$ possible for a single pixel was $\sim
10\!:\!1$ (see Ayers 1993 for a more thorough discussion).  Since a low
dispersion spectrum typically had a full width perpendicular to the
dispersion of about 3 pixels, the maximum $S/N$ of a {\em spectral pixel}
was $\sim \sqrt{3}\times10$ or $\simeq 17\!:\!1$.  Furthermore, $\sim 3$
spectral pixels make up a {\em spectral resolution element}, so the
maximum $S/N$ for a {\em single spectral element} was $\sim \sqrt{3}\times 
\sqrt{3}\times 10$, or $\simeq 30\!:\!1$. However, such high $S/N$ rarely 
occurred over a large spectral range since the limited dynamic range of the 
detector made it necessary to under expose some pixels in order to avoid 
saturating others.  Nevertheless, when multiple exposures of the same
object are averaged, a $S/N \sim 30\!:\!1$ should be attainable over 
broad spectral regions.  Therefore, our goal is to reduce the systematic 
effects in the {\it IUE}\/ data to a similar level, i.e., $\sim 3\%$.

Once an exposure was obtained, the {\it IUE}\/ cameras had to be read and
then prepared for the next exposure.  This ``read-prep'' operation 
typically required about 20 minutes.  This was often far longer than the 
actual exposure times and several of the observing strategies described 
below were intended to minimize the impact of the read-prep overhead time.

Finally, the {\it IUE}\/ satellite locked onto its target using its
Fine Error Sensor (FES).  In this mode, the satellite could point to a
fixed position on the sky with an accuracy of 0.25 arc sec.

\subsection{The observing modes}

{\it IUE}\/ spectra were obtained in one of 4 observing modes, each of
which had its own unique advantages.  These 4 observing modes were:

\noindent {\bf Large Aperture:} This mode simply involved obtaining
spectra by centering the object in the large aperture and was the
primary observing mode.  Since the large aperture was $\sim 3$ times
larger than the image of a point source produced by the {\it IUE}\/
optics, the $0.25$'' tracking capability of {\it IUE}\/ was not
critical for large aperture exposures, and the photometric quality of
these data is the best.

\noindent {\bf Small Aperture:}  In this mode, the star was centered in
the small aperture.  Since the image was considerably larger than the
aperture, the exact amount of flux entering the detector was highly
dependent upon the pointing.  Consequently, the overall photometric
quality of these data is very poor, varying by as much as a factor of
two.  Nevertheless, small aperture could be valuable for several
reasons.  First, low dispersion large and small aperture spectra were
well-separated on the detector and so both could be recorded without an
intervening read-prep.  Thus the overhead requirements and poor dynamic
range of the detectors could be circumvented to some degree by exposing
the large and small aperture spectra to different signal levels and
later combining the extracted data.  (Small aperture data would be
scaled to the photometric level defined by large aperture using data in
mutually well-exposed wavelength regions.)  Second, small aperture
spectra had slightly better spectral resolution, due to the smaller
image on the detector.  Third, sometimes it was necessary to use the
small aperture in order to isolate objects in crowded fields.

\noindent {\bf Trailed:}  In this mode, the star was allowed to drift
across the large aperture, in the cross-dispersion direction, during an
exposure.  This produced a widened spectrum which had 2 advantages. 
First, since nearly 4 times as many pixels were exposed, trailed
spectra have nearly twice the $S/N$ of large aperture spectra with the
same read-prep time.  Second, since more pixels contribute to a single
resolution element, trailed spectra are less sensitive to localized
detector irregularities (fixed pattern noise).  On the other hand, the
photometric accuracy of trailed spectra is somewhat inferior to that of
large aperture spectra because the exposure time depends on the exact
trajectory of the object through the aperture and the exact drift rate. 
Further, during the maneuver, the  spacecraft had to rely on its gyros for 
stability and could not take advantage of the FES feedback.  As a result of 
these effects, trailed spectra typically have the best relative photometric 
accuracy, but their overall flux level is slightly less accurate than large 
aperture spectra.

\noindent {\bf Multiple exposures:} In this mode, the star was placed
at 2 or 3 distinct locations perpendicular to the dispersion in the
large aperture.  The net result is a widened spectrum, although it is
not uniformly exposed (hence, this mode was often referred to as the
pseudo-trail mode).  Since FES tracking was in effect throughout the
exposures, the photometric accuracy of these spectra are probably
better than for trailed spectra, but the standard stars were not
observed often enough in this mode to verify this supposition.

For the purposes of this paper, 2 important points emerge from the
preceding discussion.
\begin{enumerate}

\item Each observing mode exposed different portions of the detector.
Therefore, the temporal and thermal behavior and absolute calibration
of each mode must be considered separately.  Unfortunately, we lack the
data to do this for the pseudo-trail mode and it will have to be
calibrated from the other modes.

\item Errors in {\it IUE}\/ spectra typically contain two distinct
components: {\em point-to-point} (or relative) errors which are important
for measuring spectral features, and {\em scaling errors}, which affect
the overall level of the spectra and are important in fitting models or in
concatenating {\it IUE}\/ spectra with each other or with optical 
photometry.  The relative magnitude of these 2 types of error depends upon 
the observing mode.  Furthermore, the scaling errors (which originate from
pointing and focus inaccuracies) can sometimes be quite large.

\end{enumerate}

There are additional effects which can have a strong influence on {\it
IUE}\/ spectra.  For instance, the particle background rate could
sometimes become rather large, reducing the $S/N$ of spectra of even
bright objects to only a few.  There were also non-random effects which
included microphonics noise (referred to as ``PINGS'') and minor frame
telemetry drop outs, which typically have a strong affect on a localized
portion of the spectrum.  See Garhart et al. (1997) for a more complete
discussion of these and other effects.

\subsection{Data selection}

Table 2 contains information on the {\it IUE}\/ NEWSIPS data used for this
analysis.  We submitted a request to the NSSDC for all of the available 
NEWSIPS low resolution spectra for the stars listed in Table 1 at the end of 
1998 March.  All but a few spectra were available at that time.  The 
delivered data were then screened as follows:

\begin{enumerate}

\item  The ranges of exposure times were restricted depending on the
object, camera, observing mode, and application.  In Table 2, $t_{min}$
and $t_{max}$ give the lower and upper limits, respectively, on acceptable
exposure times.

\item The LWR sample was restricted to the time period when the LWR was
the default long wavelength camera (1978 -- 1984).

\item  Outliers were rejected.  These were defined as follows:  The mean 
flux for each spectrum was determined over a pre-specified wavelength 
interval ($1400 - 1700$ \AA\ for the SWP and $2400 - 2800$ \AA\ for the LWR 
and LWP).  The sample mean and RMS scatter ($\sigma$) were determined for 
these mean fluxes.  If the absolute value of the difference between the mean 
flux for a given spectrum and the sample mean differed by more than 
$3\sigma$, then the spectrum was rejected.  This criterion was applied 
iteratively until no additional spectra were rejected.

\item  A few LWRs were rejected ``by hand'' because, although they passed
the outliers criteria, their shapes were distinctly peculiar. All of these
were observations of HD 60753 and most of them were affected by ``pings''
(data drop outs) or extreme background levels.

\end{enumerate}

The number of spectra which survived the screening process outlined above
is listed in the last column of Table 2 for each star and observing mode. 
Within each acceptable spectrum, data points with $\nu$-Flag values (see Garhart 
et al.\ 1997) equal to 0 (no known problem), $-128$ (on the positively 
extrapolated ITF) or $-256$ (on the negatively extrapolated ITF) were given 
weights of unity and all other points were assigned zero weight.

Finally, we note that there are two distinct ITFs for LWR\ spectra (see 
Garhart et al. 1997) and these result in slightly different wavelength
scales.  Further, ITF A (identified as LWR83R94 in the MXLO headers) results 
in spectra with 563 data points while ITF B (LWR83R96 in the headers) 
spectra have 562 points.  We used the wavelength scale from ITF\ A for all 
of the LWR spectra.  This ignores a difference between the 2 wavelength 
scales which increases linearly from 0.2 \AA\ at 1950 \AA\ to 1.66 \AA\ at 
3150 \AA\ (the longest wavelength we calibrate).  However, since even the 
largest deviation is smaller than the sampling interval (2.67 \AA) and much 
smaller than a resolution element ($\sim 7$ \AA\ at the longer wavelengths), 
we felt that interpolating the data from one grid to the other was 
unwarranted.  Therefore, we simply accept the minor systematic error which 
arises from adopting a common wavelength scale for all of the spectra.

\section{Mathematical description of the analysis}\label{anal}

To bring the NEWSIPS data onto a common scale with the FOS data, we must 
first remove its time and $T\!H\!D\!A$ dependencies.  We do this by fitting
the dependencies and then applying the results to correct the data for the 
systematics.   We then derive the transformation between the corrected data 
and the FOS system.   In this section we provide a mathematical outline of 
the problem.  We describe its application to the data in \S \ref{results}.

\subsection{The corrections}

We wish to analyze $\left\{i=1,\ldots,M\right\}$ standard stars observed
at different wavelengths and times to determine the time degradation and
$T\!H\!D\!A$ dependence of the instrumental response.  In the analysis, we 
adopt the following model of the temporal, $t$, and $T\!H\!D\!A$, $T$, 
dependence for the $i^{th}$ standard:

\begin{equation}
\label{fluxes}f(\lambda ,t,T)_i=f(\lambda,t_0,T_0)_ig(\lambda,t-t_0)
h(\lambda,T-T_0)
\end{equation}
where $g(\lambda,t-t_0)$ and $h(\lambda,T,T_0)$ are assumed to be universal 
multiplicative functions which describe the time and $T\!H\!D\!A$ 
dependencies of the instrumental response at $\lambda$ and are equal to 1 
at $(t,T)=(t_0,T_0)$.  The fact that we have written the $t$ and $T$
dependencies as separate functions implicitly assumes that the {\em form} 
of the $T\!H\!D\!A$ dependence does not change with time and that the 
temporal dependence is the same for all values of $T\!H\!D\!A$.  

Taking logarithms linearizes the problem, {\it viz.},

\begin{eqnarray}
\label{logs}\log f(\lambda,t,T)_i = & \log f(\lambda,t_0,T_0)_i 
+\log g(\lambda,t-t_0)+ \nonumber \\
& \log h(T-T_0)
\end{eqnarray}
A simple form for the functions $\log g(\lambda,t-t_0)$ and $\log 
h(\lambda,T-T_0)$ which satisfy our assumptions is
\begin{eqnarray}
\label{poly}\log g(\lambda,t-t_0)= \sum_{k=1}^{K} a(\lambda )_k(t-t_0)^k \\
\log h(\lambda,T-T_0)= \sum_{l=1}^{L} b(\lambda )_k(T-T_0)^l
\end{eqnarray}
i.e., $K$ and $L^{th}$ order polynomials.

It is possible to fit the data set for each standard star individually
using the previous equation, thereby obtaining the two sets of
coefficients, $\left\{a(\lambda)_k\right\}$ and $\left\{b(\lambda)_l
\right\}$, and the flux at the fiducial values, $\log f(\lambda,t_0)_i$, 
for each standard star separately.  However, we would like to fit the data 
of the $M$ standards {\em simultaneously}, thereby determining a universal
estimate of the coefficients, utilizing all of the available data.  To 
accomplish this, we  first concatenate the data into a single data set.  
If there are $\left\{m=1,\ldots,M\right\}$ standard stars, each with $N_m$ 
observations at times $\left\{t_n| n=1,\cdots, N_m\right\}$, at each 
wavelength $\lambda$, then the concatenated series $\left\{y(\lambda,t)
\right\}$ is defined as,

\begin{eqnarray}
\label{y} \left\{y(\lambda,t,T)\right\} \equiv & \{
\log f(\lambda, t_1)_1, \ldots, \log f(\lambda,t_{N_1})_1, \nonumber \\
& \log f(\lambda,t_1)_2,\ldots, \log f(\lambda,t_{N_2})_2,  \nonumber \\
& \ldots, \log f(\lambda,t_1)_K, \ldots,  \nonumber \\
& \log f(\lambda,t_{N_M})_{M}\}
\end{eqnarray}
The temporal and $T\!H\!D\!A$ dependence of the combined data set at each 
wavelength is then fit with a standard linear regression model of the form

\begin{eqnarray}
\label{fit} y(\lambda,t,T) = & 
  \sum_{m=1}^{M} \log f(\lambda, t_0)_m {\bf X}_{0m} + \nonumber \\
& \sum_{k=1}^{K} a(\lambda)_k {\bf X}_k  + \sum_{l=1}^{L} b(\lambda)_l {\bf Y}_l
\end{eqnarray}
where the ${\bf X}_{0m}$ are ``box-car'' functions which are either 0 or 1,
depending on whether the data refer to the $m^{th}$ star, the ${\bf X}_k$ 
are polynomials of the form $(t-t_0)^k$, and the ${\bf Y}_k$ are 
polynomials of the form $(T-T_0)^k$, where the $t$ and $T$ are the time
and $T\!H\!D\!A$ corresponding to the particular term.

As pointed out in \S \ref{data}, the major source of error in the spectra
is often an overall scaling factor due to inexact centering of the object
in the aperture or slight trailing errors.  To suppress this effect, we
normalized the spectra by their mean flux over a wavelength band
$\lambda_1 <\lambda <\lambda_2$.  These normalized spectra are denoted as
$r(\lambda,t,T)$.  The $r(\lambda,t,T)$ are fit at each wavelength by 
equation \ref{fit}, and then the normalization constants are fit 
independently the same way.  As a result, we determine 3 sets of 
coefficients, the $\{a(\lambda)_k\}$ and $\{b(\lambda)_l\}$ in equation 
\ref{fit} (except now they apply to the $r(\lambda,t,T)$) and a set 
$\{a_{0k};b_{0l}\}$, which fit the level of the flux in the standard band, 
relative to its value at $t=t_0$, $T=T_0$. Consequently, to correct the flux 
of an object observed at time $t$ with $T\!H\!D\!A =T$, one must divide the 
observed flux by the function:

\begin{eqnarray}
\label{corr}g(\lambda,t-t_0)h(\lambda,-T_0) = & \prod\limits_{k=1}^{K} 10^{
[a(\lambda)_k+a_{0k}](t-t_0)^k}\times \nonumber \\
& \prod\limits_{l=1}^{L} 10^{[b(\lambda)_l+b_{0l}](T-T_0)^l}
\end{eqnarray}
The result is how the spectrum would have appeared if it had been obtained 
at $t = t_0$ with $T\!H\!D\!A = T_0$.  

Finally, since there are not enough pseudo-trail spectra to perform an
independent calibration, this case is treated differently, and discussed in 
\S \ref{p-trail}.  

\subsection{Flux scale transformation}

Once the temporal and thermal corrections are determined, the transformation 
to the FOS flux scale is relatively straightforward.  It is simply a mean 
of the ratios of the FOS spectra of the ``flux standards'' (rebinned to {\it 
IUE}\/ resolution) to their mean {\it IUE}\/ spectra.  

\section{Application of the analysis}\label{results}

In this section we provide the details of the general analysis outlined in
the previous section, as applied to the data for the program stars listed
in Table 1.

\subsection{Temporal and $T\!H\!D\!A$ corrections}

In performing the fits, we used sixth degree polynomials in both $t$ and
$T$ for the $g(\lambda,t)$ and $h(\lambda,T)$.  The time, $t$, was expressed 
in Julian years, and $t_0=2445000/365.25$ (1 Feb. 1 1982).  The fiducial 
$T\!H\!D\!A$ value, $T_0$, was set to 9 for the SWP and LWP data and 13 for 
the LWR.  Due to a paucity of data at extreme $T\!H\!D\!A$ values, 
$T\!H\!D\!A$s of SWP and LWP data less than $T_{min} = 6$ were set equal to 
6 and values greater than $T_{max} = 13$ were set equal to 13.  The same 
process was used for the LWR data except with $T_{min} = 11$ and $T_{max} = 
16$.

For normalizing the spectra to obtain $r(\lambda, t, T)$, we used 150 data 
points in the range $2399.69 < \lambda < 2796.43$ \AA\ for the LWP and LWR 
and 179 points in the range $1400.35 < \lambda < 1698.74$ \AA\ for the SWP.

The stars HD 60753, BD$+28^{\circ} 4211$ and BD$+75^{\circ}375$ were used
as primary standards for the corrections because they have the
largest number of spectra, and these span the entire lifetime of {\it
IUE}.  BD$+33^{\circ} 2642$ and HD 93521 have the next largest number of
spectra (roughly half of any one of the primary standards), and excellent
temporal coverage.  These objects will be used to verify the results
derived from the 3 standards (see \S \ref{verify}).

\begin{figure}[h]
\figurenum{5a}
\centerline{\hbox{
\epsfxsize=5truein
\epsffile{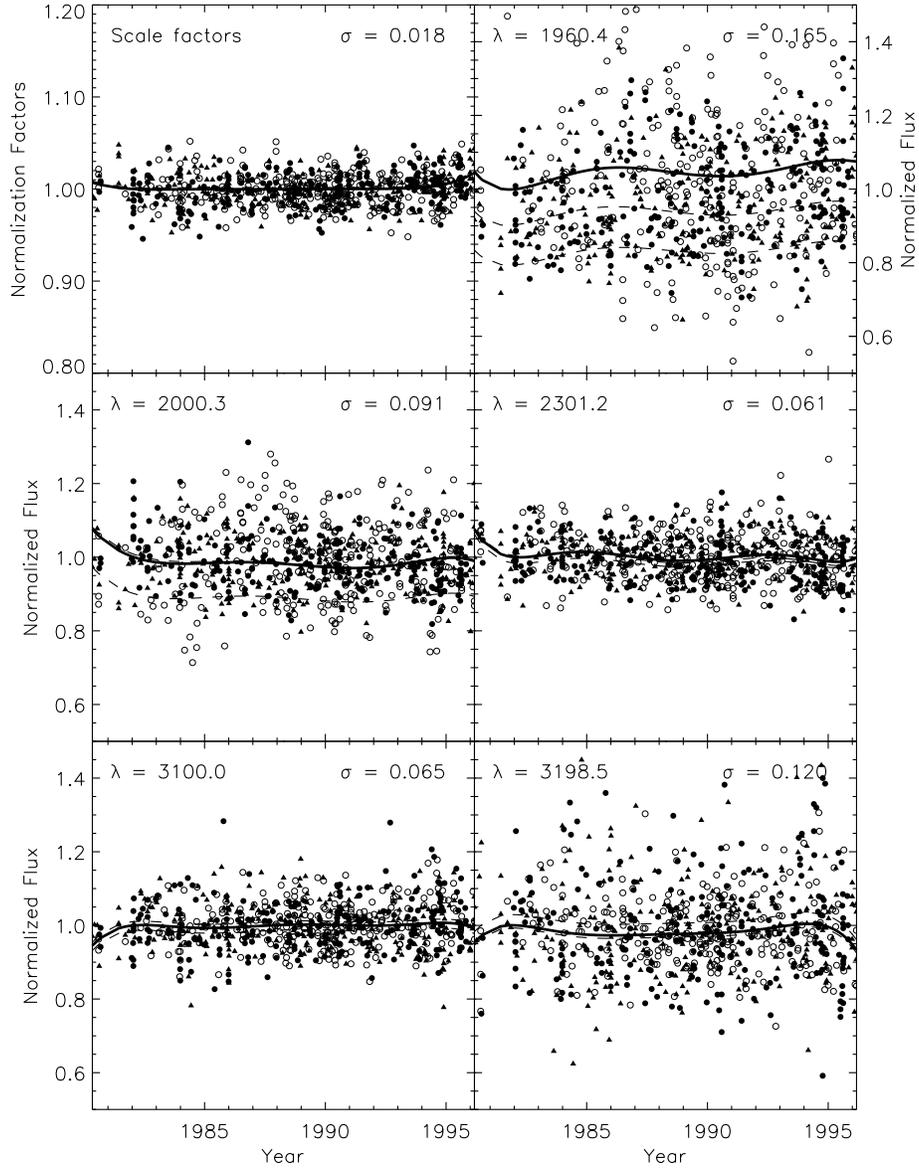}
}}
\caption{ 
Examples of fits of the model to the time dependent systematics of the LWP 
data.  The fits shown are for $T\!H\!D\!A = T\!H\!D\!A_0$, with dashed
curves indicating $\pm0.75$ of the full range in $T\!H\!D\!A$. The upper 
left panel shows the fit to the relative scale factors, and the next 5 
panels show fits to the data at selected wavelengths (listed on the 
figure).  The wavelengths are the means of 3 {\it IUE}\/ adjoining 
wavelength points to reduce the overall scatter.  Each panel also lists the 
standard deviation of the 3 channel mean fit or the scale factors, 
$\sigma$.  Data from the different stars are keyed as follows: HD 60753 -- 
open circles, BD$+28^\circ 4211$ -- filled circles, BD$+75^\circ 375$ -- 
filled triangles.}
\label{tfits}
\end{figure}

\begin{figure}[h]
\figurenum{5b}
\centerline{\hbox{
\epsfxsize=5truein
\epsffile{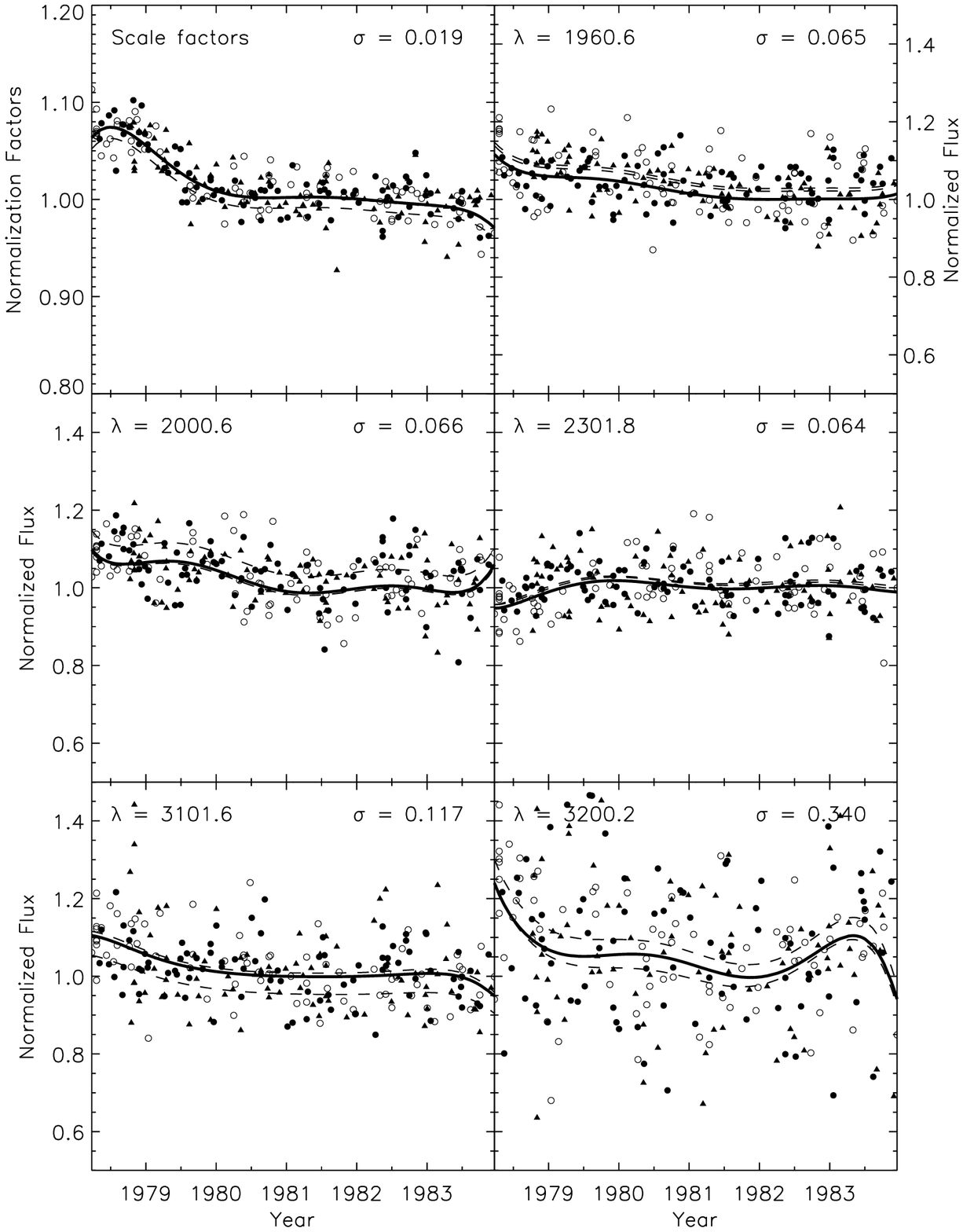}
}}
\caption{Same as Figure \ref{tfits}, for LWR data.}
\label{tfits_b}
\end{figure}

\begin{figure}[h]
\figurenum{5c}
\centerline{\hbox{
\epsfxsize=5truein
\epsffile{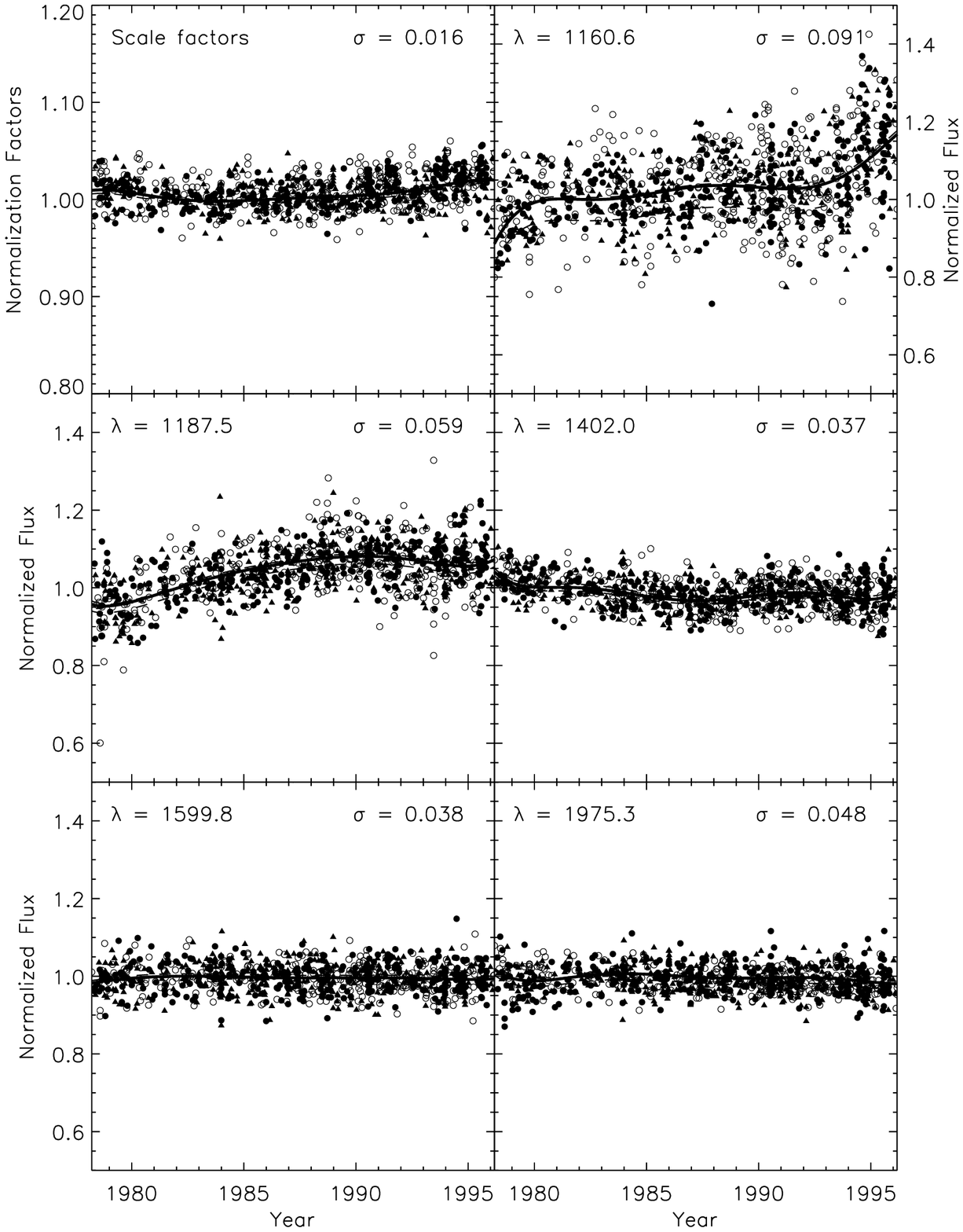}
}}
\caption{Same as Figure \ref{tfits}, for SWP data.}
\label{tfits_c}
\end{figure}

Figures \ref{tfits}-c give examples of the fits to the time trends in
the relative scale factors and at the specific wavelengths for the 3 {\it
IUE}\/ cameras.  The solid curves are the fits for $T\!H\!D\!A = T_0$, and 
the dashed curves are for the cases $T -T_0 = 0.75 (T_{max}-T_0)$ and 
$T -T_0 = 0.75 (T_{min} -T_0)$.  The fits for the extreme $T\!H\!D\!A$ 
values usually parallel the $T=T_0$ curve and typically represent a much 
smaller effect (as expected from our discussion in \S \ref{intro}).  Data 
from the 3 standards used to determine the fits are depicted by different 
symbols.  Each of the individual wavelength plots shown are actually the 
means of 3 adjoining wavelength points, to reduce the overall noise. It is 
clear that the data for the 3 stars are completely interspersed and that 
the solution is consistent with all three. Several other aspects of the 
plots are of interest.
\begin{enumerate}

\item There are no major time systematics in the LWP data (Fig.\
\ref{tfits}).

\item Data longward of 3100 \AA\ become very unreliable in both of the long
wavelength cameras (Figs.\ \ref{tfits}-b).

\item The signal-to-noise of data shortward of 2000 \AA\ is very poor in 
the LWP data, but relatively good in the LWR (Figs.\ \ref{tfits}-b).

\item The large time dependent systematic in the scaling of the LWR data
is clearly demonstrated (Fig.\ \ref{tfits_b}).

\item All but the shortest wavelengths of the SWP data have comparable
signal-to-noise (Fig.\ \ref{tfits_c}).

\item Strong systematic effects are present in the short wavelength SWP
data (Fig.\ \ref{tfits_c}).

\item The SWP data are generally of higher quality (recall that these are
all comparable exposures).  Both the scale factors and the individual
wavelength fits have smaller dispersions in the SWP data.

\end{enumerate}

\subsection{Flux transformations}

Once the spectra are corrected to their fiducial time and $T\!H\!D\!A$ 
values, the transformation to the FOS system is straightforward.  The stars 
used to determine the flux transformation were BD$+28^{\circ}4211$, 
BD$+75^{\circ}375$, BD$+33^{\circ}2642$ and G191B2B.  Both FOS and high 
quality {\it IUE}\/ data are available for each of these.  HD 60753 was not 
observed with FOS, but will provide a powerful verification of the flux
calibration.

The FOS data were first smoothed to the {\it IUE}\/ resolution using the 
data provided by Garhart et al. (1997).  These spectra were then sampled 
onto the {\it IUE}\/ grid.  There is, however, one complication.  In order 
to make the sharp He {\sc i} features located throughout the long wavelength 
{\it IUE}\/ spectra of BD$+75^\circ375$ cancel with their counterparts in 
the FOS spectra, it was necessary to adjust the wavelength scale of the long 
wavelength cameras.  Since experience has given us considerably more 
confidence in the FOS calibrations, we adopted the FOS wavelength scale and 
derived a set of adjustments for the {\it IUE}\/ scale. The measured 
differences are listed in Table \ref{wave_table}.  In practice, we applied a 
spline interpolation between these points.

\begin{figure}[h]
\figurenum{6a}
\centerline{\hbox{
\epsfxsize=5truein
\epsffile{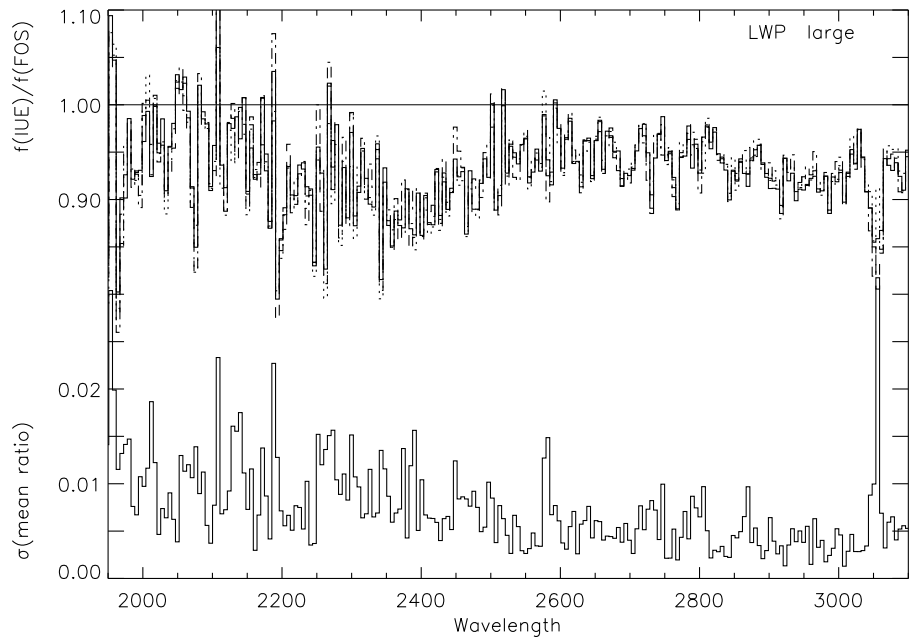}
}}
\caption{Ratios of fully corrected mean LWP large aperture spectra divided 
by FOS spectra for the 4 standard stars used to derive the absolute flux 
transformation (see Table \ref{stars_table}).  The {\it IUE}\/ wavelength 
scale was adjusted to agree with the FOS scale prior to the division.  Two
point binning was applied to the data for display.  The individual ratios 
are depicted by different curves: solid for BD$+28^\circ 4211$, dotted for 
BD$+75^\circ 375$, dashed for BD$+33^\circ 2642$, and dot-dashed for 
G191B2B.  The standard deviation of the weighted mean of the 4 curves is 
shown at the bottom.}
\label{3ratios}
\end{figure}

Figures \ref{3ratios}-c show the ratios of the completely corrected large
aperture data (with the wavelength corrections applied to the long 
wavelength data) divided by the corresponding FOS data.  Curves from the 4 
primary standards, BD$+28^\circ 4211$, BD$+75^\circ 375$, BD$+33^\circ 2642$ 
and G191B2B are depicted by different line styles.  The mean curve, used for 
the calibration, was formed by first adjusting all of the ratios to the
sample mean value across the same wavelength bands described above
and then determining a weighted mean ratio, where the weighting factors 
were just the number of observations that entered each ratio.  The standard
deviation of the weighted mean ratio was also calculated and is shown at 
the bottom of each plot.  The excellent agreement of the different 
curves emphasizes the reality of the structure, including the large 
point-to-point structure. The RMS dispersion shows that the overall 
internal agreement of the calibration curves is $\sim 1\%$ -- well within 
our goal.  The trailed and small aperture ratios have similar scatters.  It 
is interesting that the feature referred to as the 1515 \AA\ feature by 
Garhart et al.\ (1997) is clearly present.  We shall return to this point 
later.

As expected, the corrections are very similar to the curves shown in Figure 
\ref{ratios0}.  In addition to a general gray offset of $\sim 5\%$, there is 
also structure present at the $\sim 10\%$ level in all of the camera.

\subsection{Pseudo-trail spectra}\label{p-trail}

As mentioned in \S \ref{data}, the pseudo-trailed (p-trailed hereafter) 
spectra present a special problem since there are not enough of them to 
perform a thorough analysis of their time and $T\!H\!D\!A$ systematics.  
Table \ref{spectra_table} shows that there are only 18 LWP, 11 LWR and 35 
SWP p-trail spectra for the standards.  These include spectra with both 2 
and 3 exposures in the large aperture.  These subsets do not expose the 
same pixels in exactly the same way and there is no a priori reason to 
assume their corrections will be similar.  However, we are forced to assume 
that they are, since we lack the data to determine otherwise.  This 
situation is unfortunate, since in spite of the paucity of p-trail data for 
the standards, it was a popular observing mode, and there are many p-trail 
spectra in the archive.

Due to the lack of data, we had to adopted the following approach for 
calibration of the p-trail spectra.  We assume that we know the intrinsic 
flux distributions for the p-trail spectra from either FOS spectra (if 
available), or mean values of fully corrected {\it IUE}\/ large aperture 
data, transformed to the FOS system.  Each p-trail spectrum was then divided 
by its corresponding FOS or mean {\it IUE}\/ large aperture spectrum to 
produce a set of normalized spectra whose mean value should be unity.  We 
then corrected the normalized p-trail spectra with both the large aperture 
and trailed temporal and $T\!H\!D\!A$ corrections and {\it IUE}--to--FOS 
calibration and compared the results.  

We found that the large aperture corrections and calibration performed best 
in all cases; removing all obvious trends from the data, and reducing the 
overall scatter.  They also produced a mean which was uniformly close to 
unity.  This result was somewhat surprising, since the p-trails expose a 
wide swath of pixels, and one might expect their properties to be more 
similar to trailed spectra.  However, it is possible that adjacency effects 
due to ``beam pulling'' dominate.  If so, the multi-peaked cross-dispersion 
structure of the p-trail spectra may make their properties most similar to 
a single large aperture spectrum.  

In any event, the RMS scatter of the normalized p-trail spectra (corrected 
by the large aperture relationships) from unity is $\sim 1\%$ over most of 
the usable range of the SWP and LWP, but there are regions where systematic 
deviations of $\sim 3\%$ may be present.  For the LWR, the the scatter is
uniformly $\sim 3\%$, but this is largely due to the overall poor
photometric quality of the available LWR p-trails.

\subsection{Special wavelength regions}

There are wavelength regions for each camera where either the intrinsic
data or our corrections algorithms are not well-defined.  The wavelength
extremes over which the corrections can be applied were determining by
examining plots such as those shown in Figures \ref{tfits} -- \ref{tfits_c}.  
These plots show that data for the longest wavelengths of the long 
wavelength cameras are poorly defined, and applying correction factors to 
these data has little meaning.  Table \ref{bad_table} lists the wavelength 
range over which our correction factors are considered reliable for each 
camera.  The factors are set to unity outside these regions.

There are also some specific wavelength regions which are problematic.  
However, most of these should not be a real concern, since they are 
flagged by the $\nu$-Flags as being poor quality data.  However, if the
correction algorithm is blindly applied, it will produce a number for such
data.  Therefore, we caution users of our correction scheme to always use
the $\nu$-Flags to eliminate problematic data points.

One region which is {\em not} flagged by the $\nu$-Flags, but is unreliable is 
the region near Ly $\alpha$ in the SWP data. The NEWSIPS spectral extraction 
uses a low order polynomial to represent the background on either side of 
the spectrum.  This approach cannot handle Geo-coronal Ly $\alpha$ emission 
which fills the aperture in long exposures.  As a result, long exposures 
will be contaminated by Ly $\alpha$ emission over a spectral region equal 
to the projected size of the aperture used to obtain the data.  These  
ranges are 1207 -- 1222 \AA\ for large aperture are trailed data and 1210 
-- 1221 \AA\ for small aperture data.  NEWSIPS low dispersion data  
cannot be trusted in these regions, and the corrections have been set to 
unity over them. 

Finally, we note that Garhart et al.\ (1997) demonstrate that the NEWSIPS SWP 
spectra show an anomaly near 1515 \AA.  However, this problem appears to be 
significantly reduced in the corrected data, and we no longer consider it to 
be a major problem.

\section{Verification}\label{verify}

We now must verify the temporal and thermal corrections and the flux
transformations derived in the previous section.  In doing so, it is
mandatory that we use only spectra that were not employed to derive the
relationships.  Since the best data sets were used to derive the
relations, we cannot expect to test the full accuracy of the results.

We begin with verification of the temporal and $T\!H\!D\!A$ corrections.  
We use BD$+33^\circ 2642$ for verification since it and HD 93521 have the
most data of the stars not used in deriving the temporal and $T\!H\!D\!A$ 
corrections.  Figure \ref{tver} compares ratios of means of BD$+33^\circ 
2642$ spectra obtained early in the mission to means of spectra taken late 
in the mission.  Each set of ratios is labeled by the camera used to obtain 
the spectra.  The dotted curves are ratios of uncorrected NEWSIPS data and 
the solid curves are ratios of NEWSIPS data corrected for temporal and 
$T\!H\!D\!A$ systematics.  The SWP and LWP ratios are 50 spectra means and 
the LWR ratios are 20 spectra means. The mean time of each spectral mean is
provided on the ordinate labels.

While (as expected from Figure \ref{highfreq} which uses far more data)
the temporal corrections make little difference in the LWP data, they have
2 effects on the LWR and SWP spectra.  In each case, the ratios of the
corrected data are closer to unity (much more so for the LWR data) and
much of the point-to-point variation is reduced in the corrected data,
demonstrating that it was not true noise, but rather systematic effects.

We now turn to verification of the flux transformation.  Figures 
\ref{fullcor} -\ref{fullcor_c} compare the fully corrected and uncorrected 
NEWSIPS data for all of our program stars.  We begin by examining specific 
improvements in the spectra of stars used to derive the transformations and 
then turn to those stars used to verify the results.

The improvements for the stars used to derive the transformation
(BD$+75^\circ 375$, BD$+28^\circ 4211$, BD$+33^\circ$ $2642$, and G191B2B)
are truly spectacular.  In particular,

\begin{enumerate}

\item Reduction of point-to-point ``noise'' is most noticeable in the long
wavelength cameras.  In particular, the He {\sc i} lines in BD$+28^\circ
4211$ and BD$+75^\circ 375$ are much more distinct.  In fact, these
lines are barely visible in the uncorrected NEWSIPS LWR spectra of
BD$+75^\circ  375$, but are obvious in the corrected spectra.

\item The 1515 \AA\ artifact (Garhart et al. 1997) is clearly present in
the NEWSIPS spectra of BD$+28^\circ4211$, BD$+33^\circ2642$ and G191B2B,
but it is reduced or  completely removed in the corrected spectra.

\item Structure in the region $2200 < \lambda <  2500$ \AA\ in the LWP
spectra of G191B2B is removed.

\end{enumerate}

We must seek verification of these results in the stars which did not
enter into the derivation of the relationships.  We are at a bit of a
disadvantage here, since the white dwarfs not included in the derivations 
are not well observed and the OB stars have rather ``busy'' spectra.  
Nevertheless, the following  are clearly seen:

\begin{enumerate}

\item The point-to-point ``noise'' is clearly reduced in the corrected SWP
spectra of HD 60753 and dramatically reduced in the LWP and LWR spectra of
HD 93521 and HD 60753.

\item The 1515 \AA\ feature clearly reduced in the SWP spectra of GD 153,
GD 71 and HD 60753.  It is also reduced in HD 93521, but it is difficult
to see since it lies in a strongly blanketed region of its spectrum. 
Given the vastly different flux levels and temporal distributions of these
observations, the possibility that this artifact is completely removed by
the corrections is quite good.

\item The structure between 2200 and 2500 \AA\ in the LWP spectra of GD
153 and GD 71 is reduced in the corrected data.  Its removal is not so 
apparent in LWP spectra of HZ 43 and HZ 44 because of their higher noise 
level (see their FOS spectra).

\end{enumerate}

Finally, Figures \ref{fosiue}-\ref{fosiue_c} compare the fully corrected 
NEWSIPS data with the FOS data.  The FOS data have been degraded to match 
the {\it IUE}\/ spectral resolution.  For the 3 calibration stars with the 
most data, it is almost impossible to distinguish between the FOS and {\it
IUE}\ spectra.  It is also clear that the corrected {\it IUE}\/ spectra
the 4 stars not used in the calibrations agree their FOS counterparts
quite well.  The only exceptions are near Ly $\alpha$ in GD 153 and GD
71.  That disagreement arises because these were relatively long
exposures, so the region of Ly $\alpha$ is partly filled in by geo-coronal
emission (see \S 5.4).

We see, therefore, that the improvements provided by the new calibrations
are also present in spectra which were not used to derive the relationships.
This independent verification of our results provides
confidence in their veracity.

\section{Error analysis}\label{errors}

To quantify the significance of an observed feature or the accuracy of a 
flux level, two types of error must be evaluated: random and systematic 
errors.  Broadly speaking, the random errors are due to uncontrollable 
effects which change in an unpredictable manner from one exposure to the 
next.  They can be either point-to-point errors (e.g., photometric errors) 
or broad band errors (e.g., the scaling errors discussed in \S \ref{data}).  
An important aspect of random errors is that they can be ``averaged down'', 
i.e., the average of $N$ observations repeated under similar conditions is 
$1/\sqrt{N}$ times more accurate than any one of the observation.  On the 
other hand, systematic errors depend upon some specific factor (e.g., 
exposure level), are typically broad band, and cannot be averaged down, 
since the entire data set is subject to their influence.  This does not 
mean, however, that they cannot be over come.  The corrections we have 
derived for the temporal $T\!H\!D\!A$ systematics are examples of 
correctable systematic errors.

\subsection{random errors}

We begin by characterizing the random errors.  This will first be done in a 
qualitative manner, using the same approach adopted by NL.  However, we use 
HD 93521 as our test object because it was not used in the derivation of 
either the temporal or flux corrections, making it an unbiased data set.  
Further, there are enough observations of the star to determine whether the 
random $S/N$ truly asymptotes.  On the negative side, HD  93521 is known to 
have variable wind lines (e.g., Howarth and Reid 1993, Massa 1995), and its 
spectrum is rather ``busy'', containing considerable structure.

The $S/N$ was calculated exactly as outlined by NL.  First we summed $n$
spectra (where $n$ varies from 1 to the the total number in the sample)
drawn at random from the sample.  Next, we calculated means and standard
deviations over 4 point bins (roughly a resolution element).  These were
then converted into $S/N$ ratios and summed over specified wavelength
regions to obtain the final results.  The wavelength regions selected were
$1400 < \lambda < 1500$ \AA\ and $1650 < \lambda < 1900$ \AA\ for the SWP
and $2200 < \lambda <2900$ \AA\ for the long wavelength cameras.  These
regions are comprised of the most responsive portions of the cameras and
the SWP region avoids variable wind lines.

Figure \ref{noisecomp} shows the results of the analysis for both the
uncorrected and corrected NEWSIPS data.  While there is only a modest
improvement for the SWP data, the improvement for the long wavelength
camera data is quit dramatic.  The figure also shows that the LWR camera
is the most ``intrinsically'' noisy of the 3, with a maximum attainable
$S/N \sim 40\!:\!1$, followed by the SWP with a maximum $S/N \sim
60\!:\!1$ and the LWP being the best with a maximum $S/N \sim 80\!:\!1$. 
The figure also demonstrates that there is little to gain in summing more
than $\sim 10$ {\it IUE}\/ spectra, but up to that point, the gain is
considerable. 

We next consider a more quantitative description of the errors.  This is 
done by comparing the observed errors (the standard deviations derived from 
the repeated observations of the standard stars) to the NEWSIPS error model, 
whose results are given in the error vector in the NEWSIPS data files. 
Both of these are calculated as unweighted statistics, since the overall 
quality of the individual spectra are relatively uniform.  Only data points 
without a known problem ($\nu$-Flag = 0) or on the extrapolated ITF ($\nu$-Flag = 
-128 and -256) were included in each calculation.

Figure \ref{error_plt} shows ratios of the observed standard deviations, 
$\sigma(Obs)$, to errors derived from the NEWSIPS error models, 
$\sigma(N\!E\!W\!S\!I\!P\!S)$, for the 3 standard stars; HD 60753, BD
$+28^\circ 4211$ and BD$+75^\circ 375$.  The $\sigma(N\!E\!W\!S\!I\!P\!S)$ 
are simply the square root of the quadratic mean of the NEWSIPS error arrays 
for all the good data points at each wavelength, while the $\sigma(Obs)$ are 
the standard deviations of all the good points at each wavelength.  Each 
panel shows a different camera--observing mode combination.  There are 2 
curves in each panel.  One is the mean $\sigma(Obs)$ for the 3 standards 
using unscaled observations and the other is the mean $\sigma(Obs)$ derived 
from spectra which were rescaled to agree over the fixed wavelength bands 
described in \S \ref{results}.  For the large aperture data, these 2 curves 
are nearly indistinguishable, since the scaling errors discussed in \S 
\ref{data} are nearly negligible.  On the other hand, the two curves are 
well separated for the trailed data and very distinct for the small aperture 
data, with the unscaled observations always producing larger errors.

In every case, the NEWSIPS error model underestimates the actual errors. The 
amount is typically of order unity for the large aperture and trailed data, 
but more than a factor of 2 for the small aperture scaled data.  

Table \ref{error_table} lists the RMS scatter in the scaling factor as a 
fraction of the flux across the bands given in \S \ref{results} for each 
camera -- observing mode combination.  It also gives the mean 
$\sigma(Obs)/\sigma(N\!E\!W\!S\!I\!P\!S)$ ratio for each camera -- observing 
mode combination (with and without normalization).  These numbers will be 
useful guides when carrying out quantitative error analyses with NEWSIPS 
spectra, although it must be remembered that the ratios sometimes contain 
considerable shape, so their characterization as a single number can be an 
over simplification.  

\subsection{systematic errors}

There are two additional parameters provided in with the NEWSIPS data files
that can be used to search for systematic effects.  These are exposure
times and exposure levels.  

Since longer exposures typically have higher background counts, it is
possible that there could be systematics in the data which are related to
the exposure time used to obtain the spectrum.  However, the results of
Figures \ref{fosiue}-\ref{fosiue_c} argue against such a systematic.  The 
quality of the agreement between {\it IUE}\/ and FOS data for the stars
shown in the figure is excellent for stars with exposure times as short as 4
sec (for the SWP) and 6 sec (for the LWR and LWP) to as long as 21 minutes
for the LWR and 30 minutes for the LWP and SWP.  So we can be confident
that the data are free of exposure time dependent systematics over this 
range of exposure times.  
 
Finally, we examined the data for systematic differences between spectra
of the same star exposure to levels.  For this purpose, we used the net 
spectra in the NEWSIPS data files, which are expressed in linearized flux 
units, $F\!N$s (see Garhart et al.\ 1997). Systematic differences between 
spectra exposed to different mean $F\!N$s would indicate a problem with the 
Intensity Transfer Function (ITF) which transforms the observed counts (in 
data units, $D\!N$s), into the linearized $F\!N$s.  

To search for an ITF problem, we examined spectra of the same star obtained 
with different exposure times, making sure that saturated pixels were 
eliminated from the comparison.  Although no major ($\geq 3\%$) systematics 
were uncovered in the SWP and LWR data, the LWP spectra do contain sizable 
ITF systematics.  Figure \ref{itf} shows exposure level systematics for LWP 
spectra of BD$+28^\circ 4211$, HD~60753, and BD$+75^\circ 375$.  The plot 
shows mean NEWSIPS fluxes over the wavelength band $2350<\lambda<2400$\AA\
(the peak  of the LWP camera response) for each star.  These have been
normalized by the mean of all exposures with $200<F\!N<400$ (saturated 
pixels {\em and} pixels using an extrapolated ITF were eliminated).  
Observations of the same star with very different mean $F\!N$s result from 
different exposure times.  The fact that fluxes derived from long exposure 
times (large $F\!N$ values) are systematically larger (by about 5\%) than 
fluxes derived from low $F\!N$s indicates a problem with the LWP ITF (since
we have already ruled out systematics which depend solely on exposure 
time).  This means that a comparison of an optimal exposure and a half
optimal exposure of stars with similar energy distributions will contain 
systematic errors up $\sim 5\%$.  

Figure \ref{itf2} shows how the ITF problem can also affect the shape of an 
energy distribution.  It displays ratios of long and short exposures of  
BD$+28^\circ 4211$ (full curve), HD~60753 (dotted curve), and
BD$+75^\circ  375$ (dashed curve).  The ratios consist of mean spectra
derived from spectra whose mean $F\!N$ over the band $2650 <\lambda<2700$ 
\AA\ lie in the range $200< <F\!N> <400$, divided by means of spectra with 
$600< <F\!N> <800$ over the same band.  Data  from saturated pixels and 
pixels using an extrapolated ITF were excluded from the means.  It is clear 
from the figure how data near the peak of the camera response (between 2500 
and 2850 \AA) are systematically different by up to 5\%.  

The LWP ITF problem is not as severe as it first appears.  It means that 
comparisons of well exposed and under exposed data may have systematics on
the order of 5\%.  However, 5\% is roughly the size of random errors for 
half optimal exposures (the reason the systematic effect shows up so well in 
Figure \ref{itf2} is that hundreds of spectra went into the ratios), so the 
effect does not dominate the errors when comparing a single half optimal 
exposure to an optimal one (although it is comparable to the random errors).  
Another property that tends to suppress the effect is that the shape of the 
response curve dominates the shape of the net spectrum.  Thus, as long as 
the intrinsic shapes of the objects being compared are not too different and 
as long as the spectra have similar exposure levels, the impact of the LWP 
ITF problem should not be too bad.   However, it underscores our assertion 
that the results of the present paper strictly apply to early type stars 
only.

\section{Summary and Conclusion}\label{summary}

\begin{enumerate}

\item We have analyzed more than 4600 spectra to demonstrate that low 
dispersion NEWSIPS data contain systematic effects on the order of $10-15$\% 
and to obtain corrections for these effects.

\item Systematics were reduced to $< 3\%$ in most instances, but can be 
as large as 5\% in a few specific cases involving LWP data.  Overall, we
can hope for a $S/N \sim 30\!:\!1$ ($\sim 20\!:\!1$ for some LWP
applications) but not more.  To exceed this value one would have to 
consider recalibrating the ITFs, and even rederiving them from first
principles.

\item Nevertheless, it may be possible to surpass the 3\% level when
dealing with relative measurements of a very homogeneous data set obtained
over a relatively short period of time and under similar conditions.

\item We have derived a transformation between the corrected {\it IUE}\/ 
data and the {\it HST}\/ FOS absolute flux scale.  The magnitude of the 
transformations can be larger than 10\% at certain wavelengths.

\item We note that much of what appears to be noise in NEWSIPS spectra is 
actually the result of high frequency systematics and flux calibrations.

\item The random errors in the corrected NEWSIPS data are characterized in 
Table \ref{error_table}.

\item The pseudo trailed spectra are poorly characterized by the available 
data.  However, application of the large aperture corrections and flux 
transformation appear to reduce their systematics to about the 3\% level.

\item We emphasize that our results apply to NEWSIPS data for {\em blue 
objects}, and we cannot guarantee any broader application.

\item Finally, a set of IDL procedures which apply the results of this 
paper to NEWSIPS low dispersion spectra will be made available to the 
{\it IUE}\/ RDAF project at the Space Telescope Science Institute and will 
be available from the authors on request.

\end{enumerate}

\acknowledgments We would like to thank Michael Van Steenberg and Nancy
Oliversen for sharing their detailed knowledge of {\it IUE}\/ data and how
it is processed.  We are also thankful to Patricia Lawton and Karen Levay
for facilitating access to the data and to Joy Nichols for providing us
with the model of G191B2B used by the {\it IUE}\/ project to calibrate the
data. We also acknowledge support through NASA contract NAG5-7372 to 
Raytheon STX and Grant NAG5-7113 to Villanova University.

\newpage
\onecolumn

\begin{deluxetable}{lcrrcccc}
\tablewidth{0pc}
\tablecaption{Program Stars \label{stars_table}}
\tablehead{
\colhead{Star} & \colhead{Spectral} & \colhead{$V$} & \colhead{$B-V$} 
& \colhead{Temporal/} & \colhead{Temporal/}& \colhead{Flux}& \colhead{Flux} \\
\colhead{} & \colhead{Type} & \colhead{} & \colhead{} 
& \colhead{Thermal} & \colhead{Thermal}& \colhead{Standard}& 
\colhead{Control} \\
\colhead{} & \colhead{} & \colhead{} & \colhead{} &
\colhead{Standard} & \colhead{Control} & \colhead{} & \colhead{}}
\startdata
HD 60753   & B3 IV  &  6.69 & -0.09 & $\times$ &    --  &  --  & $\times$ \\
HD 93521   & O9.5Vn &  7.04 & -0.28 &  --    & $\times$ &  --  & $\times$ \\
BD$+28^{\circ}4211$ & sdO & 10.52 & -0.33 & $\times$ & -- & $\times$ & -- \\
BD$+75^{\circ}325$  & sdO &  9.54 & -0.37 & $\times$ & -- & $\times$ & -- \\
BD$+33^{\circ}2642$ & B2 IVp & 10.84 & -0.17 & --& $\times$ & $\times$&-- \\
G191B2B    & DA     & 11.78 & -0.34 &  --    & $\times$ & $\times$ & -- \\
GD 71      & DA     & 13.04 & -0.24 &  --    &   --     &  --  & $\times$ \\
GD 153     & DA     & 13.42 & -0.25 &  --    &   --     &  --  & $\times$ \\
HZ 43      & DA     & 12.86 & -0.10 &  --    &   --     &  --  & $\times$ \\
HZ 44      & DA     & 11.71 & -0.27 &  --    &   --     &  --  & $\times$ \\
\enddata
\end{deluxetable}

\begin{deluxetable}{lccrrr}
\tablewidth{0pc}
\tablecaption{Spectra \label{spectra_table}}
\tablehead{
\colhead{Star} & \colhead{Camera} & \colhead{Observation} & \colhead{$t_{min}$}
& \colhead{$t_{max}$} & \colhead{Number of Spectra} \\
\colhead{}          & \colhead{} & \colhead{Mode}      & \colhead{(sec)} &
\colhead{(sec)}      & \colhead{Used in Analysis}}
\startdata
HD 60753    & LWP    & lg apt    &   4   &    8  &  258   \nl
            & --     & trail     &  10   &   50  &  212   \nl
            & --     & sm apt    &   8   &   20  &   98   \nl 
            &        & p-trail   &  --   &   --  &    7   \nl \tableline
HD 60753    & LWR    & lg apt    &   4   &   20  &   79   \nl
            & --     & trail     &   8   &   45  &   77   \nl
            & --     & sm apt    &   5   &   22  &   52   \nl 
            &        & p-trail   &  --   &   --  &    2   \nl \tableline
HD 60753    & SWP    & lg apt    &   6   &   12  &  321   \nl
            & --     & trail     &  10   &   75  &  270   \nl
            & --     & sm apt    &  10   &   30  &  145   \nl 
            &        & p-trail   &  --   &   --  &   11   \nl  \tableline \tableline
HD 93521    & LWP    & lg apt    &   2   &    4  &  114   \nl
            & --     & trail     &   7   &   16  &   23   \nl
            & --     & sm apt    &   5   &   10  &   16   \nl 
            &        & p-trail   &  --   &   --  &    2   \nl \tableline
HD 93521    & LWR    & lg apt    &   2   &    4  &   68   \nl
            & --     & trail     &  12   &   20  &   11   \nl
            & --     & sm apt    &   4   &   10  &   48   \nl 
            &        & p-trail   &  --   &   --  &    1   \nl \tableline
HD 93521    & SWP    & lg apt    &   2   &  4.5  &  171   \nl
            & --     & trail     &  10   &   20  &   34   \nl
            & --     & sm apt    &   4   &    9  &   61   \nl 
            &        & p-trail   &  --   &   --  &    2   \nl \tableline \tableline
BD$+28^{\circ}4211$ & LWP    & lg apt    &  25   &   55  &  244   \nl
            & --     & trail     &  90   &  500  &   36   \nl
            & --     & sm apt    &  75   &  160  &   51   \nl
            &        & p-trail   &  --   &   --  &    4   \nl \tableline
BD$+28^{\circ}4211$ & LWR    & lg apt    &  30   &   80  &   81   \nl
            & --     & trail     & 100   &  450  &   16   \nl
            & --     & sm apt    &  40   &  190  &   46   \nl 
            & --     & p-trail   &  --   &   --  &    4   \nl   \tableline
BD$+28^{\circ}4211$ & SWP    & lg apt    &  20   &   60  &  350   \nl
            & --     & trail     &  40   &  160  &   55   \nl
            & --     & sm apt    &  20   &   90  &  100   \nl   
            & --     & p-trail   &  --   &   --  &    7   \nl   \tableline \tableline
BD$+33^{\circ}2624$  & LWP    & lg apt   &  120  &  220  &  120  \nl
            & --     & trail     &  --   &   --  &    1   \nl
            & --     & sm apt    &  300  &  600  &    4   \nl 
            & --     & p-trail   &  --   &   --  &    3   \nl  \tableline
BD$+33^{\circ}2624$  & LWR    & lg apt   &   80  &   200 &   54  \nl
            & --     & trail     &  --   &   --  &    1   \nl
            & --     & sm apt    &  200  &  500  &   13   \nl  
            & --     & p-trail   &  --   &   --  &    3   \nl  \tableline
\tablebreak
BD$+33^{\circ}2624$  & SWP    & lg apt   &  150  &  350  &  179  \nl
            & --     & trail     &  --   &   --  &    3   \nl
            & --     & sm apt    &  250  &  500  &   14   \nl 
            & --     & p-trail   &  --   &   --  &    9   \nl  \tableline \tableline
BD$+75^{\circ}375$  & LWP    & lg apt    &  10   &   30  &  248   \nl
            & --     & trail     &  40   &  180  &   50   \nl
            & --     & sm apt    &  30   &   70  &   72   \nl 
            & --     & p-trail   &  --   &   --  &    2   \nl   \tableline
BD$+75^{\circ}375$  & LWR    & lg apt    &  10   &   35  &   77   \nl
            & --     & trail     &  20   &  100  &   20   \nl
            & --     & sm apt    &  25   &   75  &   50   \nl  
            & --     & p-trail   &  --   &   --  &    1   \nl   \tableline
BD$+75^{\circ}375$  & SWP    & lg apt    &  12   &   36  &  321   \nl
            & --     & trail     &  15   &  110  &   77   \nl
            & --     & sm apt    &  15   &   45  &  116   \nl  
            & --     & p-trail   &  --   &   --  &    3   \nl   \tableline  \tableline
G191B2B     & LWP    & lg apt    &   --  &   --  &   36   \nl
            & --     & sm apt    &   --  &   --  &    2   \nl  \tableline
G191B2B     & LWR    & lg apt    &   --  &   --  &    1   \nl
            & --     & trail     &   --  &   --  &    1   \nl  \tableline
G191B2B     & SWP    & lg apt    &   --  &   --  &   34   \nl
            & --     & trail     &   --  &   --  &   14   \nl
            & --     & sm apt    &   --  &   --  &    5   \nl 
            & --     & p-trail   &  --   &   --  &    4   \nl  \tableline \tableline
GD 71       & LWP    & lg apt    &   --  &   --  &    8   \nl  \tableline 
GD 71       & LWR    & lg apt    &   --  &   --  &    1   \nl  \tableline 
GD 71       & SWP    & lg apt    &   --  &   --  &    7   \nl
            & --     & sm apt    &   --  &   --  &    2   \nl 
            & --     & p-trail   &  --   &   --  &    4   \nl  \tableline \tableline
GD 153      & LWP    & lg apt    &   --  &   --  &    8   \nl  \tableline 
GD 153      & LWR    & lg apt    &   --  &   --  &    2   \nl  \tableline 
GD 153      & SWP    & lg apt    &   --  &   --  &   10   \nl
            & --     & sm apt    &   --  &   --  &    2   \nl  \tableline \tableline
HZ 43       & LWP    & lg apt    &   --  &   --  &    1   \nl  \tableline 
HZ 43       & LWR    & lg apt    &   --  &   --  &    5   \nl
            & --     & sm apt    &   --  &   --  &    2   \nl  \tableline 
HZ 43       & SWP    & lg apt    &   --  &   --  &    9   \nl
            & --     & sm apt    &   --  &   --  &    4   \nl  
            & --     & p-trail   &  --   &   --  &    4   \nl  \tableline \tableline
HZ 44       & LWP    & lg apt    &   --  &   --  &    3   \nl  \tableline 
HZ 44       & LWR    & lg apt    &   --  &   --  &    1   \nl
HZ 44       & SWP    & lg apt    &   --  &   --  &    7   \nl  % \tableline \tableline
\enddata
\end{deluxetable}

\clearpage

\begin{deluxetable}{rr}
\tablewidth{0pc}
\tablecaption{LWP/LWR Wavelength Corrections \label{wave_table}}
\tablehead{
\colhead{$\lambda$ (\AA)} & \colhead{$\Delta \lambda$ (\AA)}}
\startdata
1800  & -2.50  \nl
1950  & -2.50  \nl
2200  & -2.00  \nl
2250  & -1.75  \nl
2300  & -1.50  \nl
2385  & -1.00  \nl
2510  & -0.40  \nl
2730  &  0.00  \nl
3500  &  0.00  \nl % \tableline
\enddata
\end{deluxetable}

\begin{deluxetable}{lc}
\tablewidth{0pc}
\tablecaption{Wavelength Ranges  \label{bad_table}}
\tablehead{
\colhead{Camera} & \colhead{Range}}
\startdata
LWP    & 1950 -- 3150 \AA \nl
LWR    & 1850 -- 3150 \AA \nl
SWP    & 1150 -- 1978 \AA \nl
\enddata
\end{deluxetable}

\begin{deluxetable}{llccc}
\tablewidth{0pc}
\tablecaption{$R\!M\!S$ errors and scale factors \label{error_table}}
\tablehead{
\colhead{Camera} & \colhead{Observing} & \colhead{Scaling error} & 
\colhead{$\sigma(Obs)/\sigma(NEWSIPS)$} & 
\colhead{$\sigma(Obs)/\sigma(NEWSIPS)$} \\
\colhead{} & \colhead{mode} & \colhead{} & \colhead{un-normalized}
& \colhead{normalized}}
\startdata
LWP    & Large Apt  &  0.020   &  1.15  &  1.12 \nl
       & Trailed    &  0.025   &  1.44  &  1.24 \nl
       & Small Apt  &  0.304   &  7.93  &  2.69 \nl \tableline
LWR    & Large Apt  &  0.038   &  1.11  &  1.09 \nl
       & Trailed    &  0.036   &  1.18  &  1.03 \nl
       & Small Apt  &  0.180   &  3.91  &  2.01 \nl \tableline
SWP    & Large Apt  &  0.018   &  1.28  &  1.23 \nl
       & Trailed    &  0.021   &  1.49  &  1.30 \nl
       & Small Apt  &  0.264   & 10.3  &  2.68 \nl \tableline
\enddata
\end{deluxetable}

\clearpage

%%%%%%% FIGURE 1 %%%%%%%%%%%%%%

%%%%%%% FIGURE 2 %%%%%%%%%%%%%%

%%%%%%% FIGURE 3 %%%%%%%%%%%%%%

%%%%%%% FIGURE 4 %%%%%%%%%%%%%%

\begin{figure}[h]
\figurenum{4b}
\centerline{\hbox{
\epsfxsize=4.5truein
\epsffile{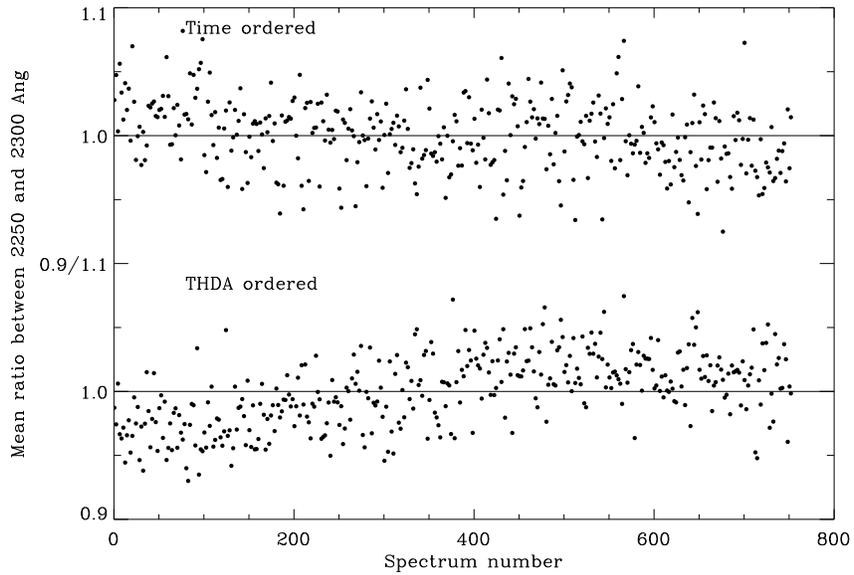}
}}
\caption{Same as \ref{thdasys}, for LWP data binned over the wavelength
range $2250 \leq \lambda \leq 2300$.  In this case, adjoining points were 
averaged to reduce the larger noise level due to the smaller wavelength 
band used to demonstrate the effect.}
\label{thdasys_b}
\end{figure}

%%%%%%% FIGURE 5 %%%%%%%%%%%%%%

%%%%%%% FIGURE 6 %%%%%%%%%%%%%%

\begin{figure}[h]
\figurenum{6b}
\centerline{\hbox{
\epsfxsize=5truein
\epsffile{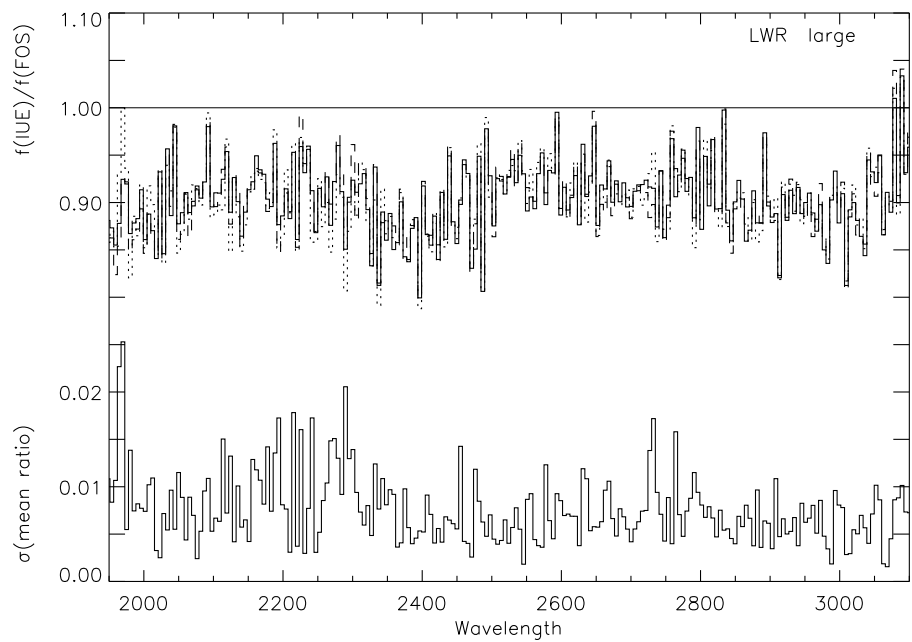}
}}
\caption{Same as Figure \ref{3ratios} for LWR data.  
G191B2B is excluded due to a paucity of data.}
\label{3ratios_b}
\end{figure}

\begin{figure}[h]
\figurenum{6c}
\centerline{\hbox{
\epsfxsize=5truein
\epsffile{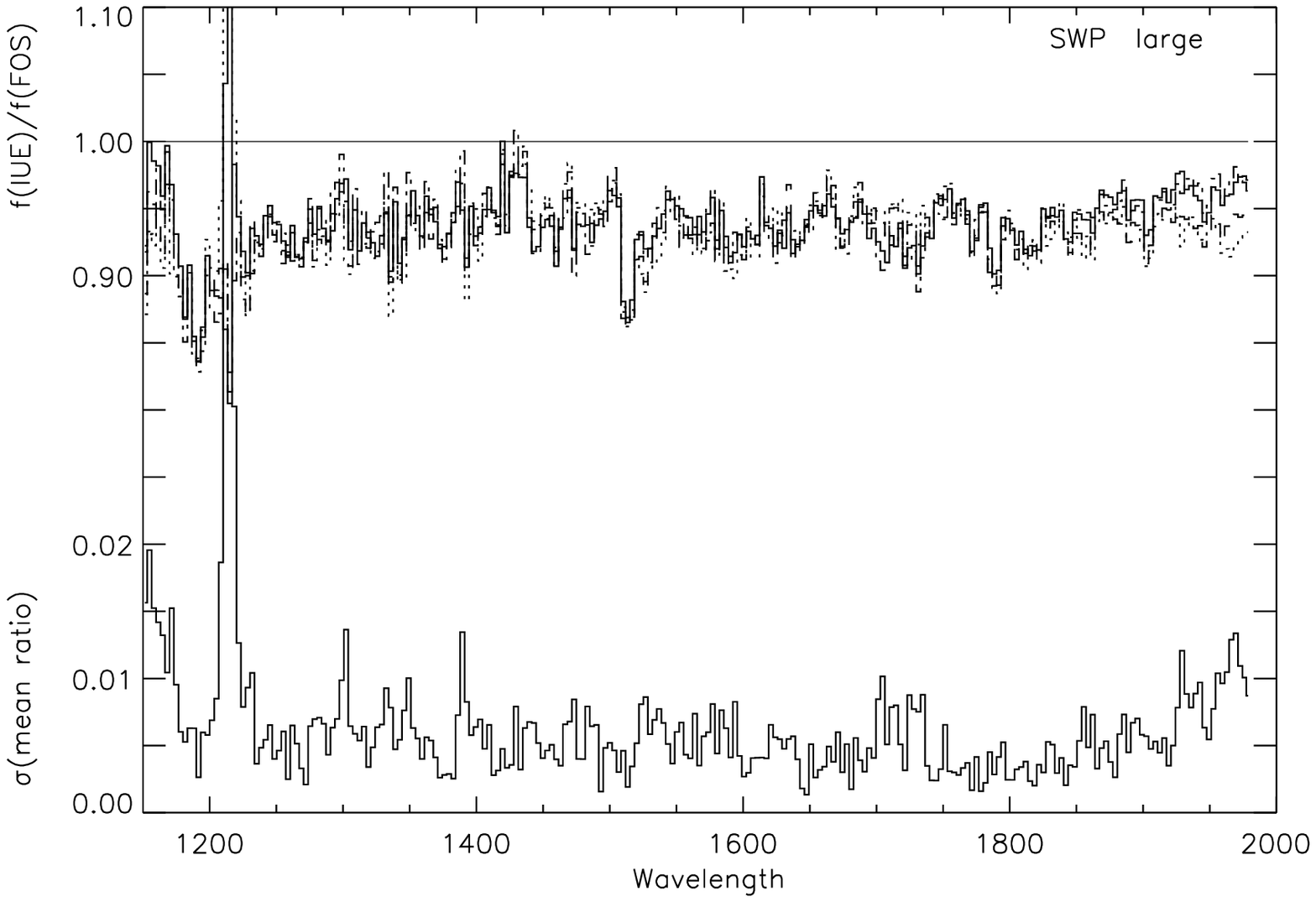}
}}
\caption{Same as Figure \ref{3ratios} for the SWP data.}
\label{3ratios_c}
\end{figure}

%%%%%%% FIGURE 7 %%%%%%%%%%%%%%
\clearpage
\setcounter{figure}{6}
\newpage
\begin{figure}[h]
\centerline{\hbox{
\epsfxsize=5truein
\epsffile{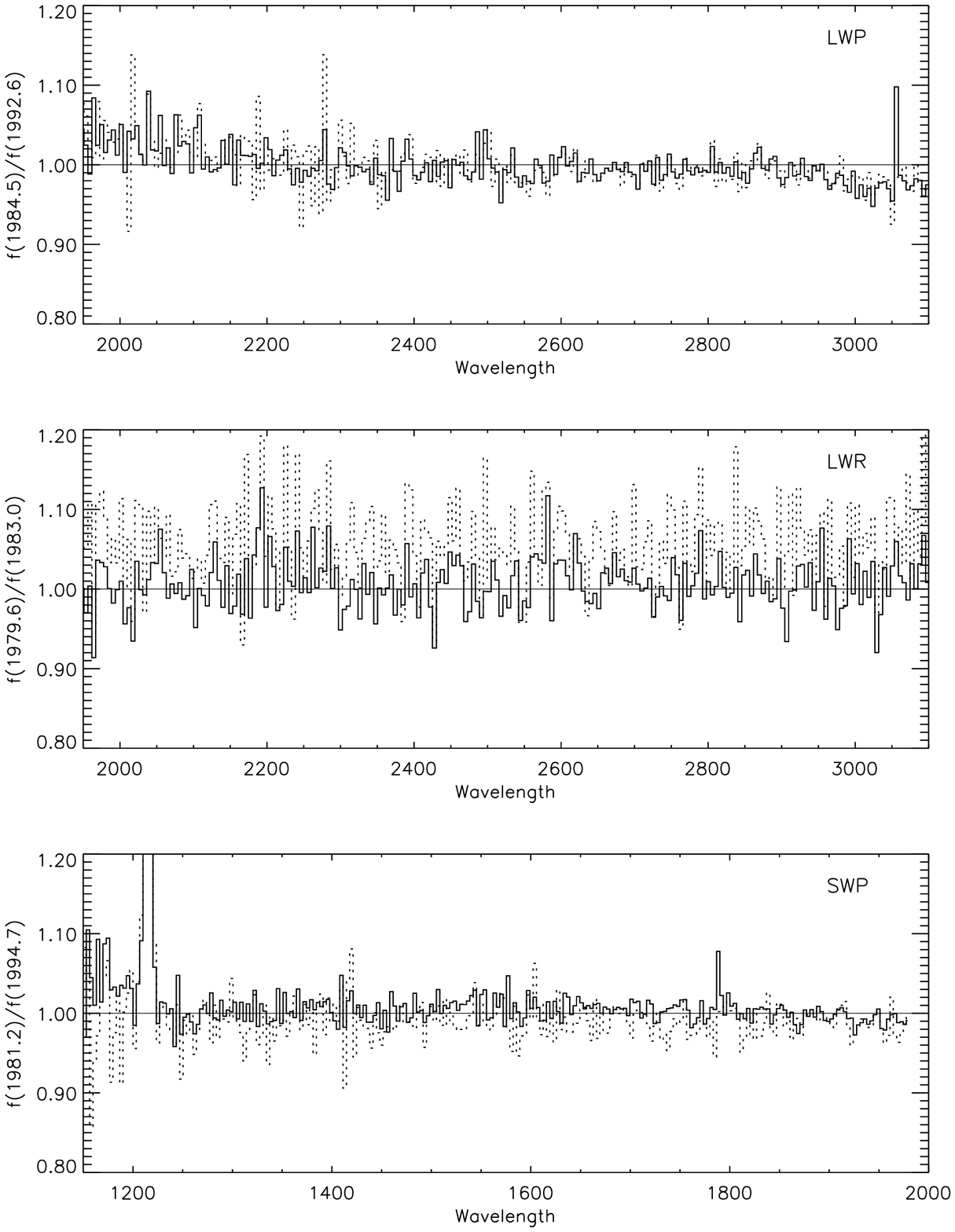}
}}
\caption{Ratios of means of spectra obtained early in the mission to means
of spectra taken late in the mission for BD$+33^\circ 2642$.  Each set of
ratios is labeled by the camera used to derive them.  The dotted curves are
ratios of uncorrected NEWSIPS data and the solid curves are ratios of
NEWSIPS data corrected for systematics.  The SWP and LWP ratios are
50 spectra means and the LWR ratios are 20 spectra means.  The mean time of
the spectra are given on the ordinate labels.}
\label{tver}
\end{figure}

%%%%%%% FIGURE 8 %%%%%%%%%%%%%%
\newpage
\begin{figure}[h]
\figurenum{8a}
\centerline{\hbox{
\epsfxsize=6.2truein
\epsffile{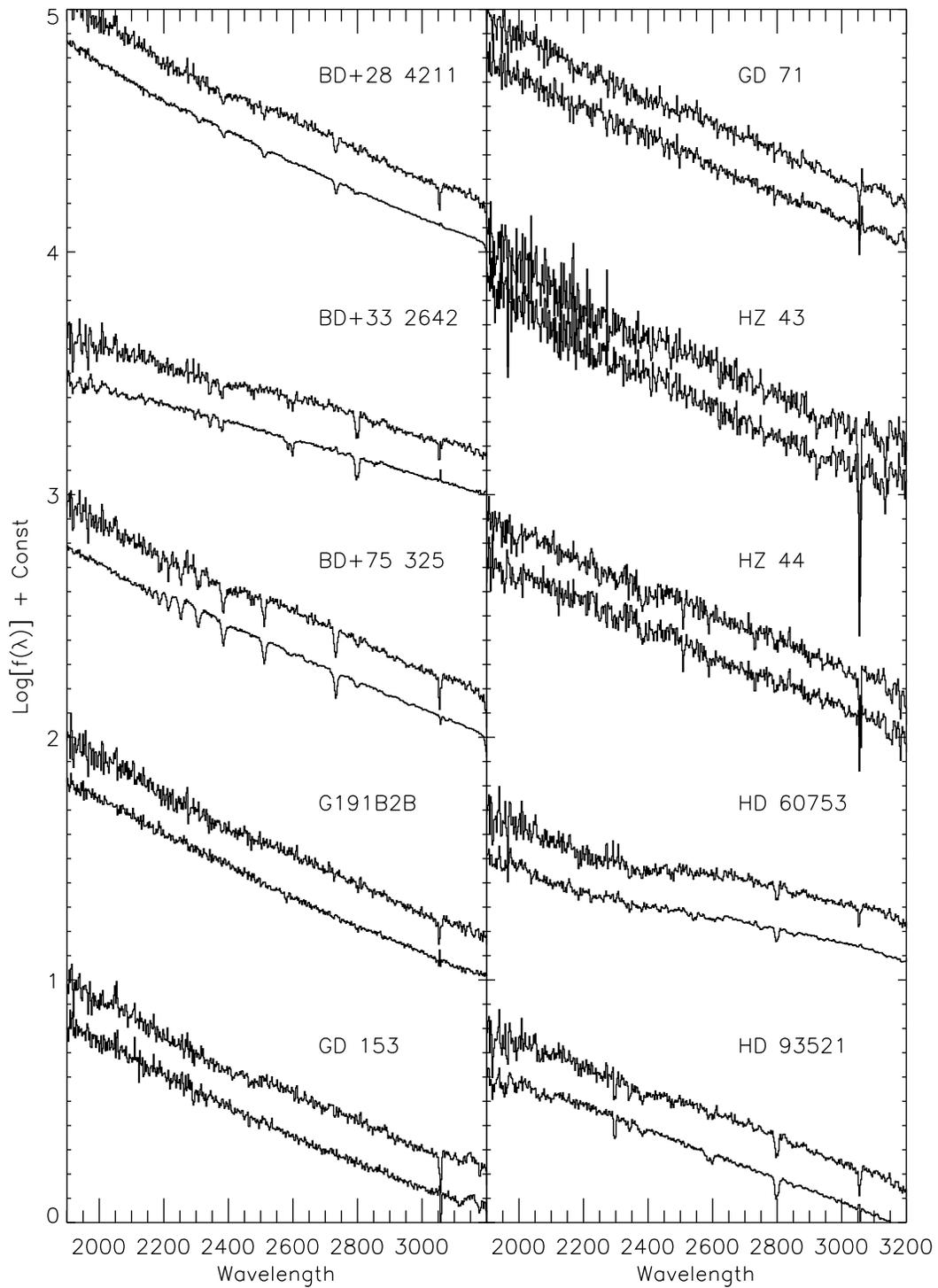}
}}
\caption{Comparison of NEWSIPS and fully corrected NEWSIPS LWP data.  Each 
pair of plots has the original, uncorrected NEWSIPS fluxes above the fully 
corrected data transformed to the FOS flux scale.}
\label{fullcor}
\end{figure}

\newpage
\begin{figure}[h]
\figurenum{8b}
\centerline{\hbox{
\epsfxsize=6.2truein
\epsffile{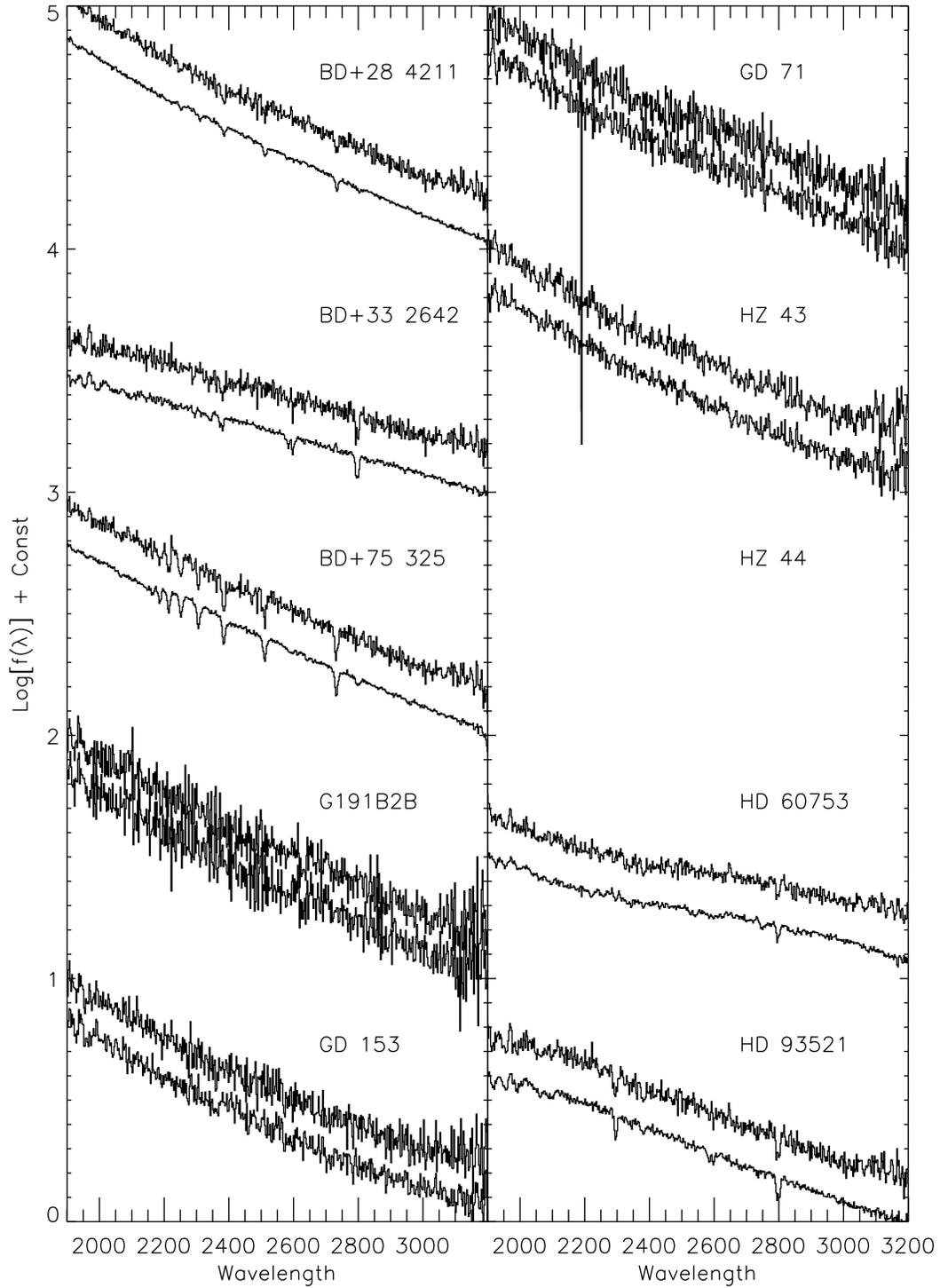}
}}
\caption{Same as Figure \ref{fullcor} for the LWR 
data.  There are no LWR large aperture data for HZ 44.}
\label{fullcor_b} 
\end{figure}

\newpage
\begin{figure}[h]
\figurenum{8c}
\centerline{\hbox{
\epsfxsize=6.2truein
\epsffile{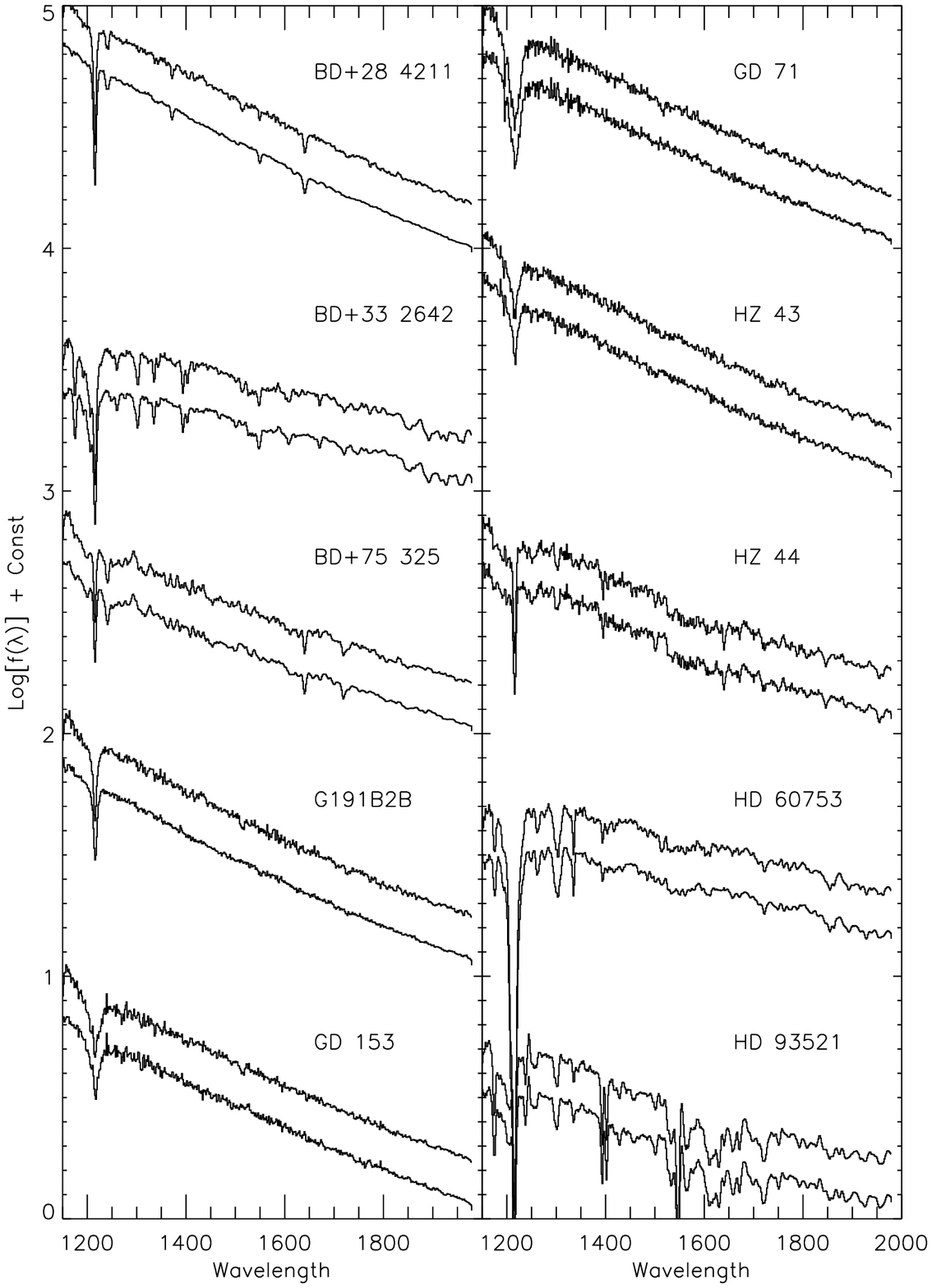}
}}
\caption{Same as Figure \ref{fullcor} for the SWP data.}
\label{fullcor_c} 
\end{figure}

%%%%%%% FIGURE 9 %%%%%%%%%%%%%%
\newpage
\begin{figure}[h]
\figurenum{9a}
\centerline{\hbox{
\epsfxsize=6.2truein
\epsffile{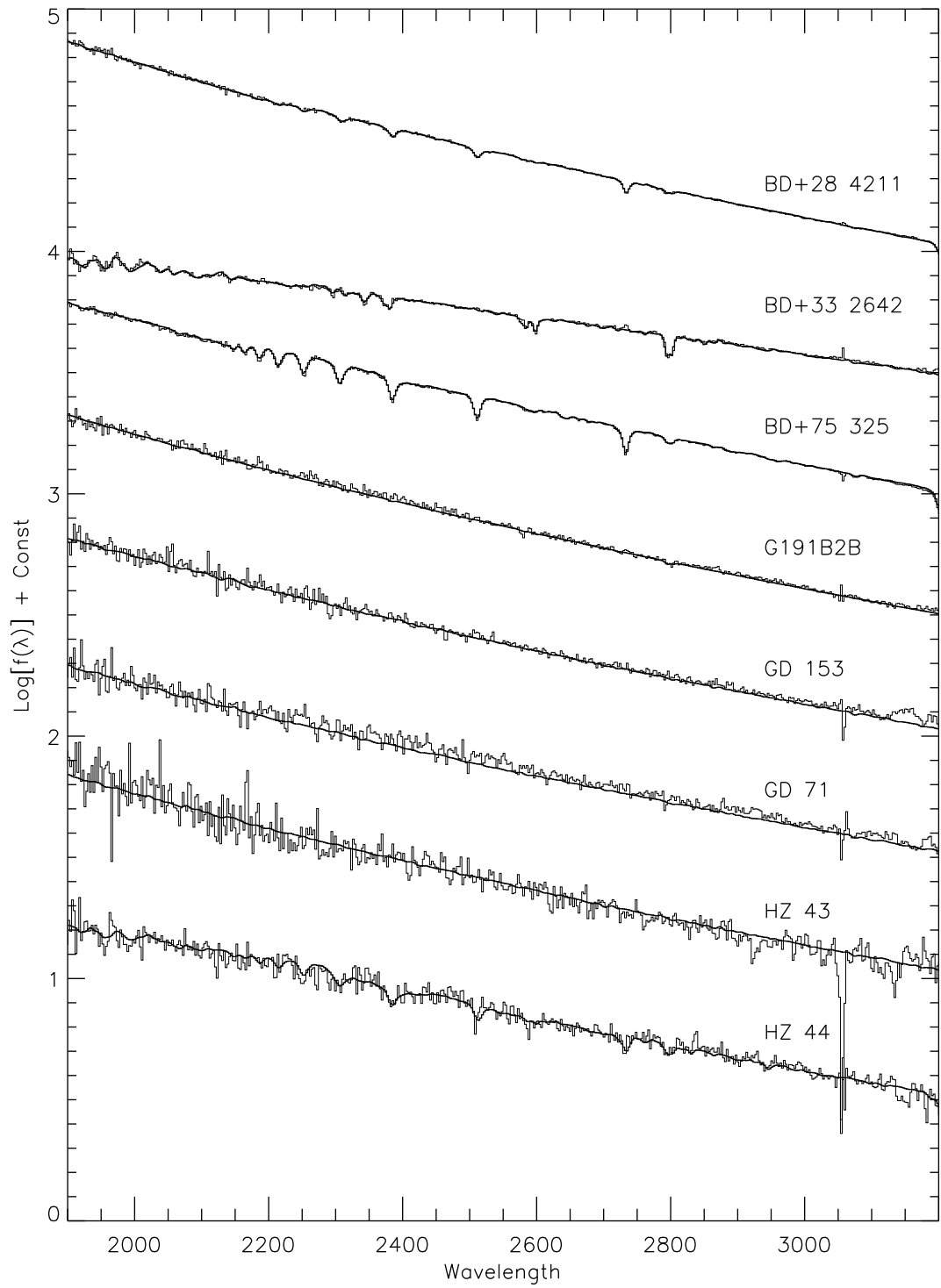}
}}
\caption{Comparison of FOS data (thick curve) and fully corrected
NEWSIPS LWP data (thin curve) for stars in common.  The FOS data have been
degraded to the {\it IUE}\/ resolution.}
\label{fosiue} 
\end{figure}

\newpage
\begin{figure}[h]
\figurenum{9b}
\centerline{\hbox{
\epsfxsize=6.2truein
\epsffile{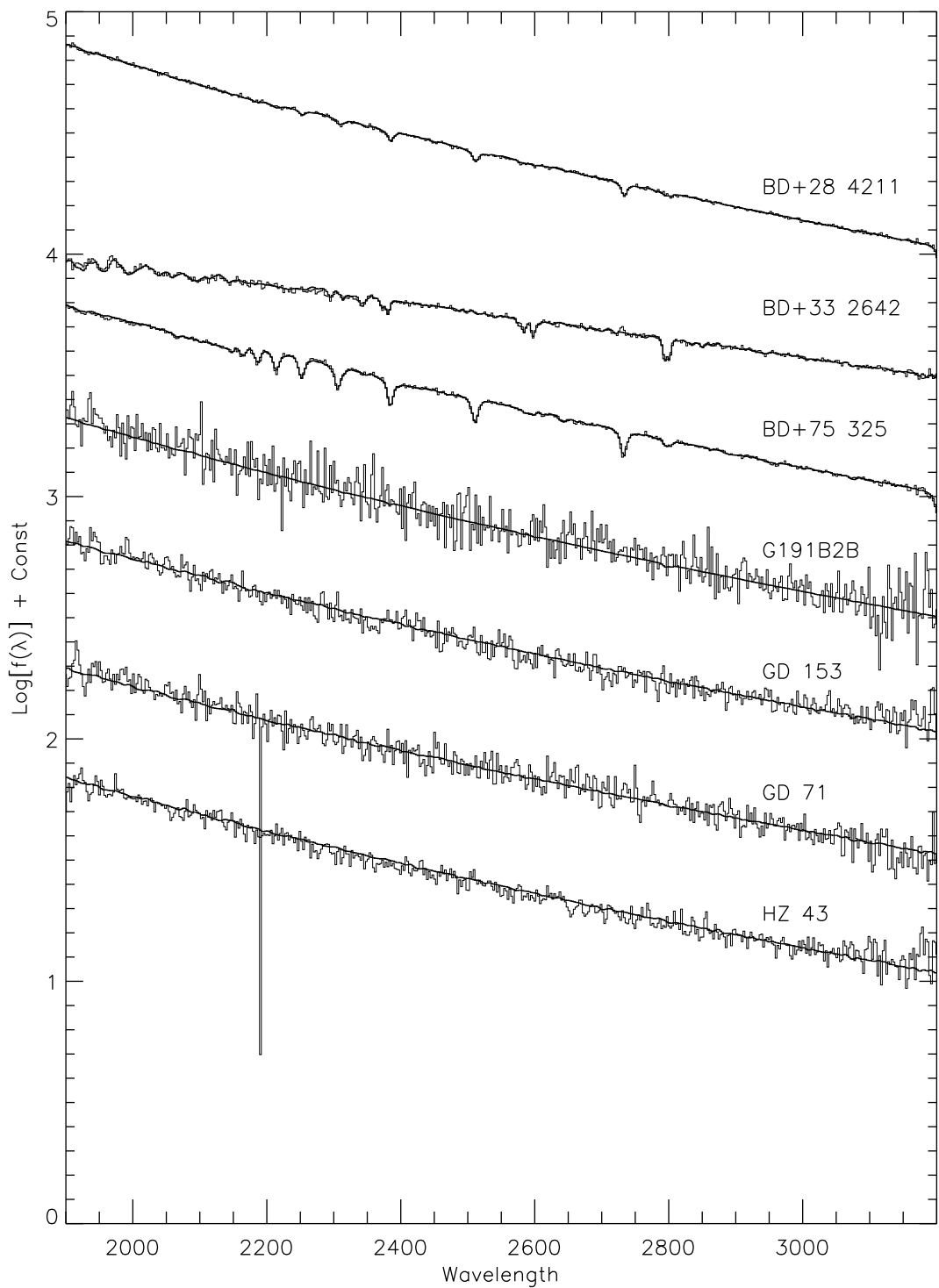}
}}
\caption{Same as Figure \ref{fosiue} for the LWR data.}
\label{fosiue_b} 
\end{figure}

\newpage
\begin{figure}[h]
\figurenum{9c}
\centerline{\hbox{
\epsfxsize=6.2truein
\epsffile{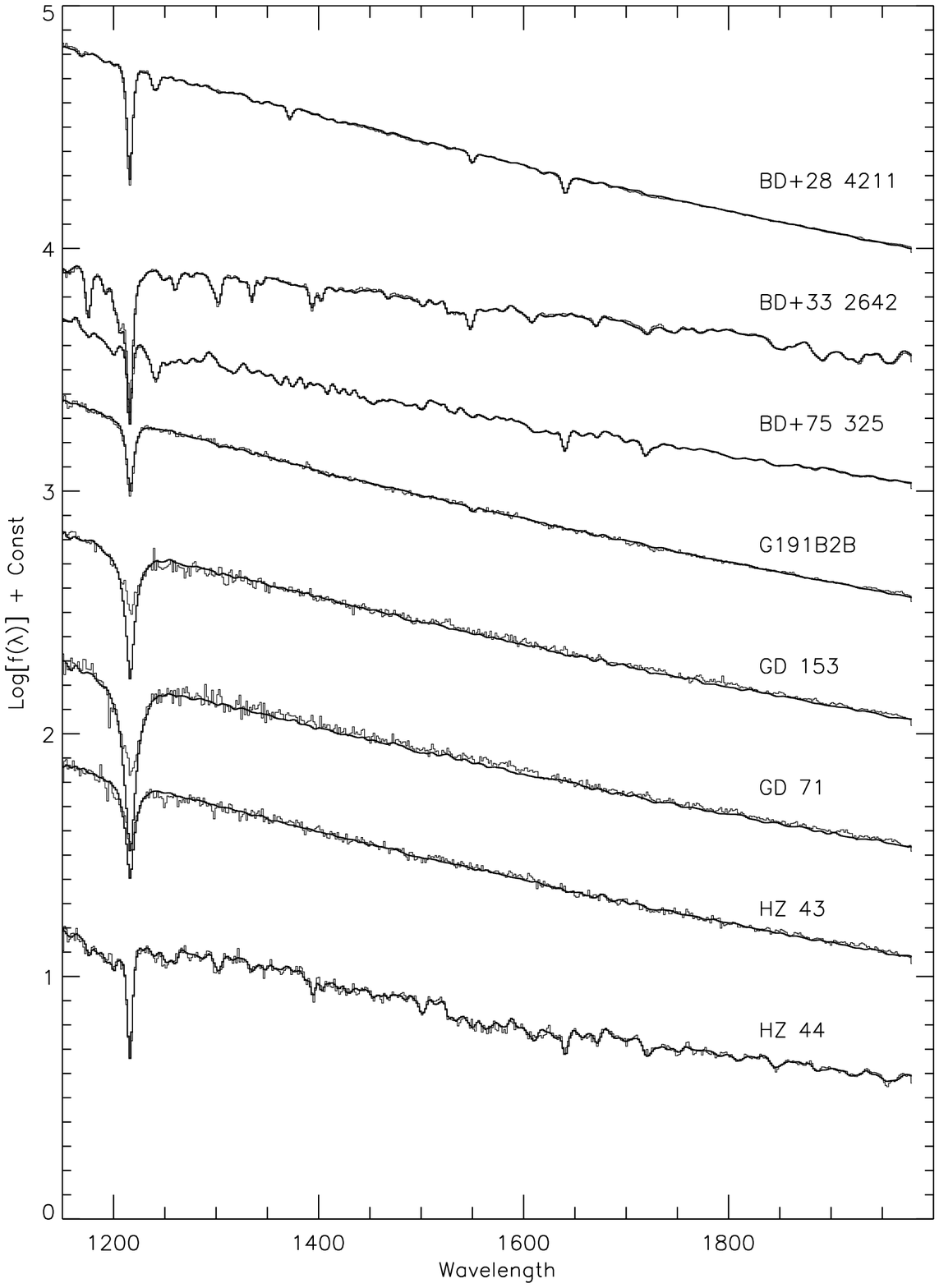}
}}
\caption{Same as Figure \ref{fosiue} for the SWP data.}
\label{fosiue_c} 
\end{figure}

\newpage
%%%%%%% FIGURE 10 %%%%%%%%%%%%%%
\setcounter{figure}{9}
\begin{figure}[h]
\centerline{\hbox{
\epsfxsize=5.0truein
\epsffile{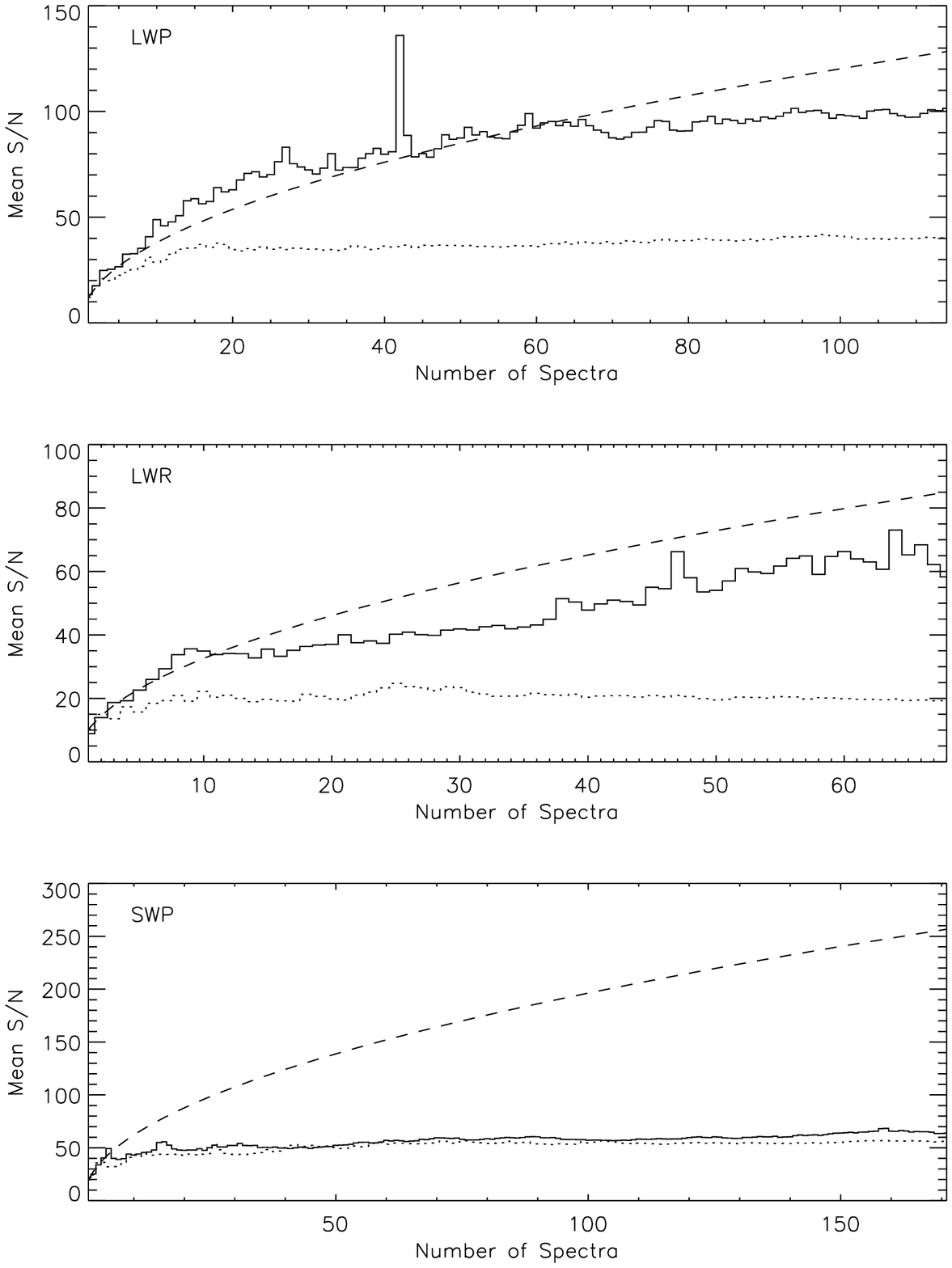}
}}
\caption{Comparison of $S/N$ values (see \S \ref{errors}) as a function of 
the number of spectra averaged to make the mean.  Large aperture spectra of 
HD 93521 were used in the calculations.  Results determined from NEWSIPS 
data are indicated as dotted curves, fully corrected NEWSIPS results are 
solid curves and the theoretical, systematic-free limits are dashed curves.  
The relevant camera is indicated in each plot.}
\label{noisecomp}
\end{figure}

%%%%%%% FIGURE 11 %%%%%%%%%%%%%%
\begin{figure}[h]
\centerline{\hbox{
\epsfxsize=5.0truein
\epsffile{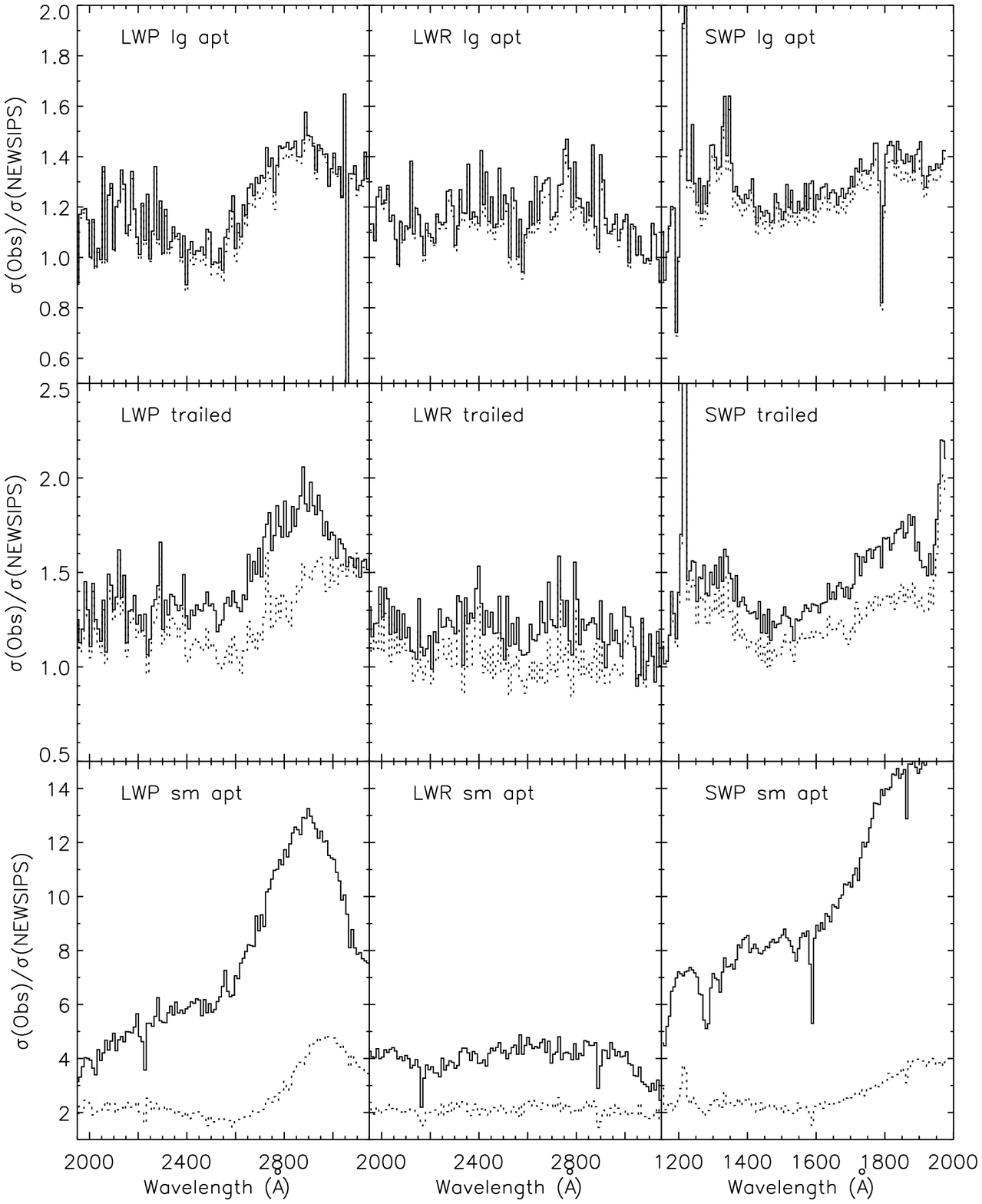}
}}
\caption{Ratio of the observed standard deviations, $\sigma(Obs)$, to 
errors derived from the NEWSIPS error models,
$\sigma(N\!E\!W\!S\!I\!P\!S)$, for the mean of the 3 standard stars; 
HD 60753, BD$+28^\circ 4211$ and BD$+75^\circ 375$.  There is one panel 
for each camera-observing mode combination, and there are 2 curves for each
star.  The solid curves are for $\sigma(Obs)$ derived from fully corrected, 
unscaled observations and the dotted curve is for $\sigma(Obs)$ derived 
from fully corrected spectra which have been rescaled to a common mean over 
a fixed wavelength band.}
\label{error_plt}
\end{figure}

%%%%%%% FIGURE 12 %%%%%%%%%%%%%%
\begin{figure}[h]
\centerline{\hbox{
\epsfxsize=4.0truein
\epsffile{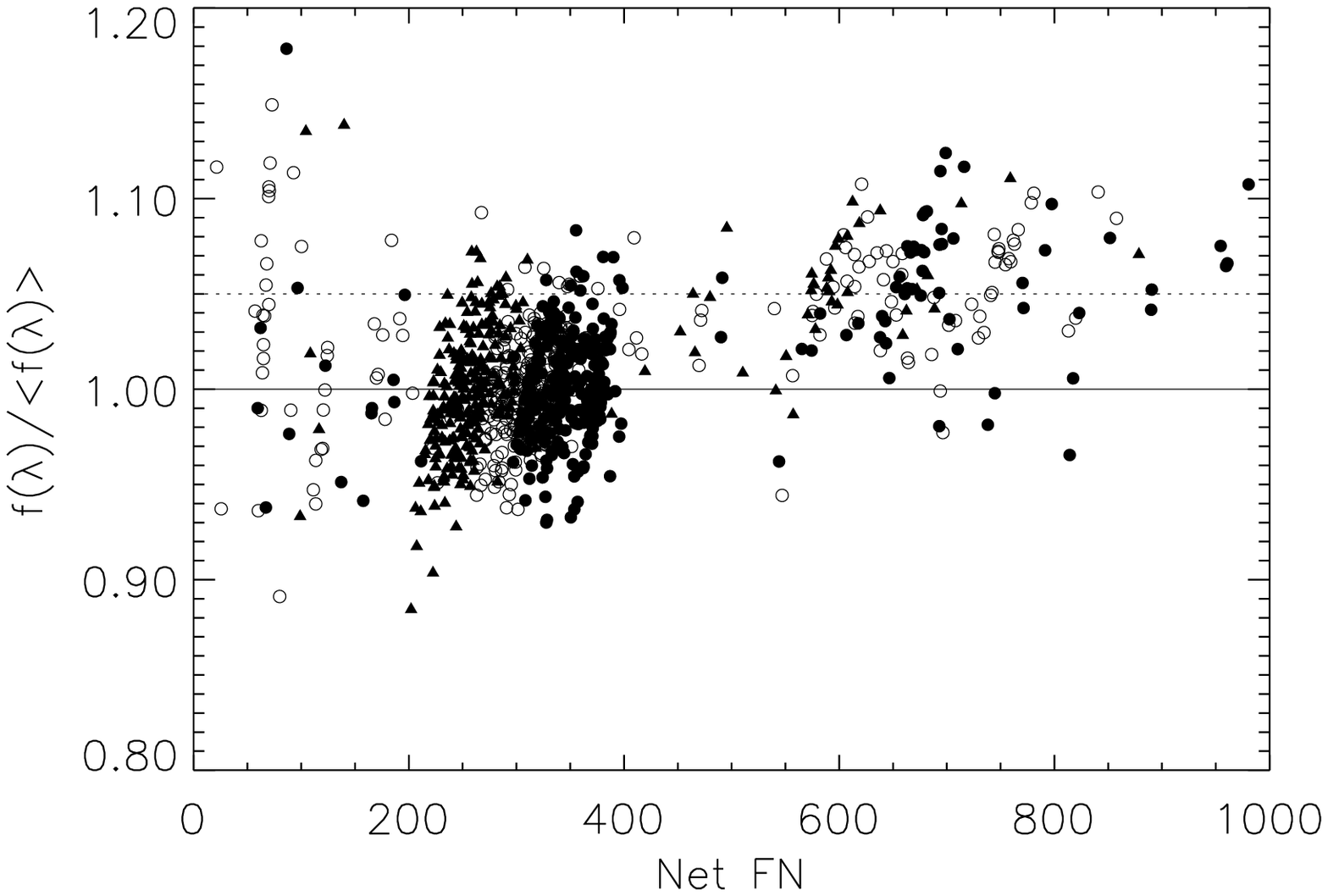}
}}
\caption{Exposure level systematics in BD$+28^\circ 4211$ (full circles),
HD~60753 (triangles), and BD$+75^\circ 375$ (open circles).  The plot shows
mean NEWSIPS flux values over the wavelength region $2350<\lambda<2400$ \AA\
for each star divided by the mean for all exposures with linearized flux 
numbers ($F\!N$) in the range $200<F\!N<400$.  Fluxes derived from
saturated pixels and pixels from an extrapolation of the ITF were not 
included.  This figure demonstrates that LWP fluxes derived from exposures 
with large $F\!N$ values are systematically larger than average, indicating 
a problem with the LWP ITF (see, \S \ref{errors}.}
\label{itf}
\end{figure}

%%%%%%% FIGURE 13 %%%%%%%%%%%%%%
\begin{figure}[h]
\centerline{\hbox{
\epsfxsize=4.0truein
\epsffile{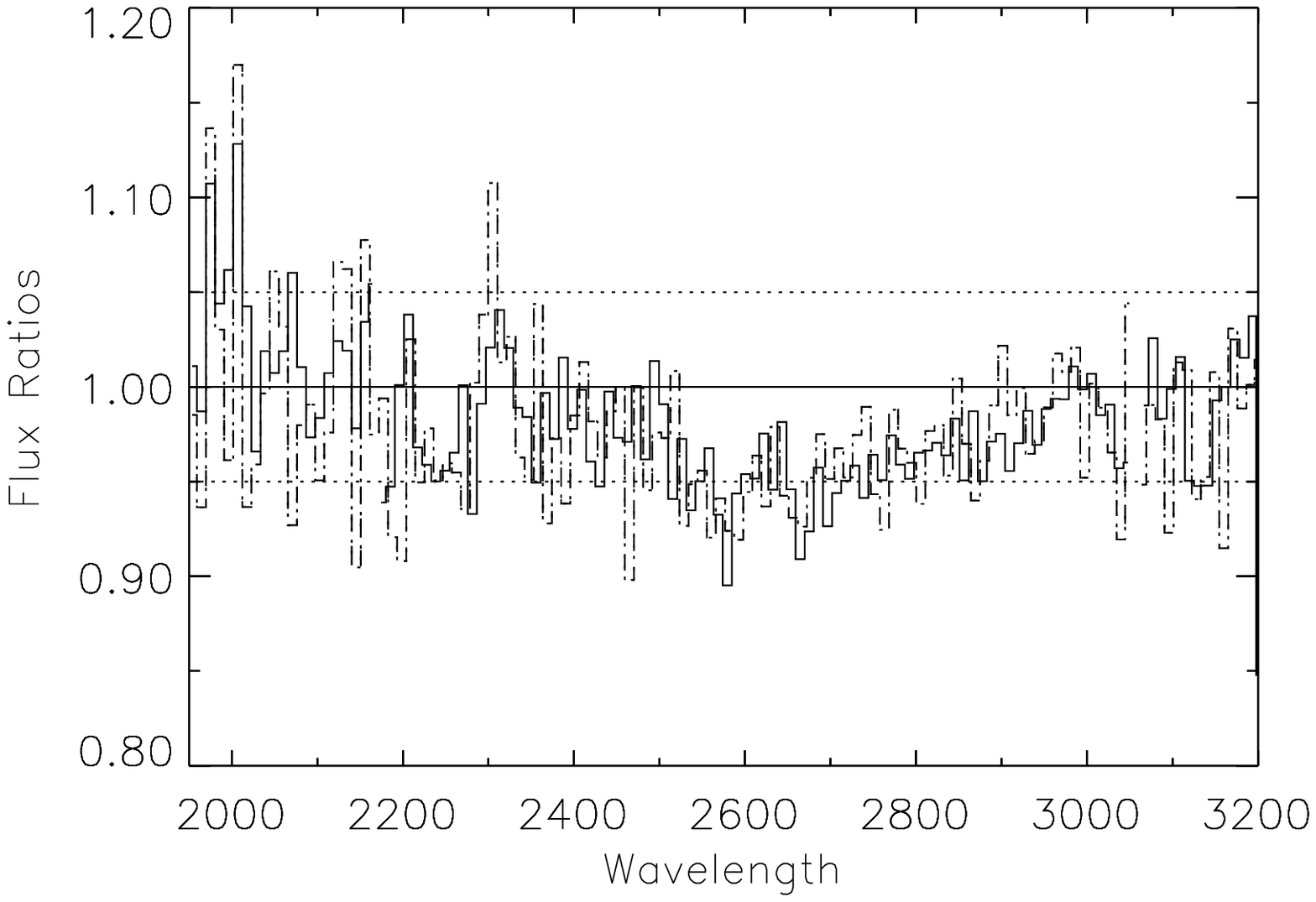}
}}
\caption{Ratios of long and short exposures for BD$+28^\circ 4211$ (full
curve), HD~60753 (dotted), and BD$+75^\circ 375$ (dashed).  The plot shows
the ratio of mean NEWSIPS fluxes for spectra selected to have $F\!N$ values in
the range $200<F\!N<400$ over the wavelength band $2650<\lambda<2700$ \AA\
divided by spectra selected to have $F\!N$ values in the range $200<F\!N<400$
over the same band.  Data from saturated pixels and pixels using an 
extrapolated ITF were excluded.}
\label{itf2}
\end{figure}
\end{document}